\documentclass[aps,preprint]{revtex4}%
\usepackage{amsfonts}
\usepackage{amsmath}
\usepackage{amssymb}
\usepackage{graphicx}%
\providecommand{\U}[1]{\protect\rule{.1in}{.1in}}

\begin{document}
\title{Holographic field theory models of dark energy in interaction with dark matter}
\author{Sandro M. R. Micheletti \footnote{smrm@fma.if.usp.br }}
\affiliation{Instituto de F\'{\i}sica, Universidade de S\~{a}o Paulo, CP 66318, 05315-970,
Sao Paulo, Brazil}

\begin{abstract}
We discuss two lagrangian interacting dark energy models in the context of the
holographic principle. The potentials of the interacting fields are
constructed. The models are compared with CMB distance information, baryonic
acoustic oscilations, lookback time and the Constitution supernovae sample.
For both models the results are consistent with a non vanishing interaction in
the dark sector of the universe - with more than three standard deviations of
confidence for one of them. Moreover, in both cases, the sign of coupling is
consistent with dark energy decaying into dark matter, alleviating the
coincidence problem.

\end{abstract}
\maketitle


\thispagestyle{empty}

\section{Introduction}

In the last years, there have been several papers where an interaction in the
dark sector of the universe is considered \cite{1} - \cite{sandro3}. A
motivation to considering the interaction is that dark energy and dark matter
will evolve coupled to each other, alleviating the coincidence problem
\cite{1}. A further motivation is that, assuming dark energy to be a field, it
would be more natural that it couples with the remaining fields of the theory,
in particular with dark matter, as it is quite a general fact that different
fields generally couple. In other words, it is reasonable to assume that there
is no symmetry preventing such a coupling between dark energy and dark matter
fields. Using a combination of several observational datasets, as supernovae
data, CMB shift parameter, BAO, etc., it has been found that the coupling
constant is small but non vanishing within at least $1\sigma$ confidence level
\cite{1} - \cite{sandro2}. In two recent works, the effect of an interaction
between dark energy and dark matter on the dynamics of galaxy clusters was
investigated through the Layser-Irvine equation, the relativistic equivalent
of virial theorem \cite{peebles}. Using galaxy cluster data, it has been shown
that a non vanishing interaction is preferred to describe the data within
several standard deviations \cite{virial}. However, in most of these papers,
the interaction term in the equation of motion is derived from
phenomenological arguments. It is interesting to obtain the interaction term
from a field theory. Some works have already taken a step in such a direction
\cite{sandro2}, \cite{sandro3}, \cite{bean}. On the other hand, scalar fields
have been largely used as candidates to dark energy. They naturally arise in
particle physics and string theory. Good reviews on this subject can be found
in \cite{copeland}. A motivation to use scalar fields as candidates to dark
energy is that their pressure can be negative, making possible to reproduce
the recent period of accelerated expansion of the universe. For example, for
canonical scalar field, the equation of state parameter varies between $1$ and
$-1$. Dark energy modeled as a canonical scalar field is called quintessence
and was investigated, for example, in \cite{bean}, \cite{canonico}. For
tachyon scalar field, the equation of state is always negative. The tachyon
field has been studied in recent years in the context of string theory, as a
low energy effective theory of D-branes and open strings \cite{sen}. Tachyon
field as dark energy was studied, for example, in \cite{sandro2},
\cite{taquions}, \cite{taqholo}. The first question about scalar fields
concerns the choice of the potential. Common choices are the power law and the
exponential potentials. However, these choices are in fact arbitrary. In
principle, any other form for the potential which leads to recent accelerated
expansion would be acceptable.

On the other hand, it is possible that a complete understanding of the nature
of dark energy will only be possible within a quantum gravity theory context.
Although results for quantum gravity are still missing, or at least premature,
it is possible to introduce, phenomenologically, some of its principles in a
model of dark energy. Recently combinations of quintessence, quintom and
tachyon models with holographic dark energy had been proposed - in
\cite{escalarholo}, \cite{quintomholo} and \cite{taqholo}, respectively.
Specifically, by imposing that the energy density of the scalar field must
match the holographic dark energy density, namely $\rho_{\Lambda}=3c^{2}%
M_{Pl}^{2}L^{-2}$, where $c$ is a numerical constant and $L$ is the infrared
cutoff, it was demonstrated that the equation of motion of fields for the
non-interacting case reproduces the equation of motion for holographic dark
energy. In fact, to impose that the energy density of scalar field must match
the holographic dark energy density corresponds to specify its
potential\textbf{.} This can be seen as a physical criterion to choose the
potential. Here, we generalize this idea for two kinds of interacting scalar fields.

\section{The models}

We consider the general action%
\begin{equation}
S=\int d^{4}x\sqrt{-g}\left\{  \frac{M_{Pl}^{2}}{2}R+\mathcal{L}_{\varphi
}\left(  x\right)  +\frac{i}{2}[\bar{\Psi}\gamma^{\mu}\nabla_{\mu}\Psi
-\bar{\Psi}\overleftarrow{\nabla}_{\mu}\gamma^{\mu}\Psi]-(M-\beta\varphi
)\bar{\Psi}\Psi+\sum\limits_{j}\mathcal{L}_{j}\left(  x\right)  \right\}
\label{action}%
\end{equation}
where $M_{Pl}\equiv\left(  8\pi G\right)  ^{-1/2}$ is the reduced Planck mass,
$R$ is the curvature scalar, $\mathcal{L}_{\varphi}\left(  x\right)  $ is,
unless of the coupling term, the lagrangian density for the scalar field,
which we will identify with dark energy, $\Psi$ is a massive fermionic field,
which we will identify with dark matter, $\beta$ is the dimensionless coupling
constant and $\sum\limits_{j}\mathcal{L}_{j}\left(  x\right)  $ contains the
lagrangian densities for the ramaining fields. Notice that, in this work, we
will only consider an interaction of dark energy with dark matter. If there
was a coupling between the scalar field and baryonic matter, the corresponding
coupling constant $\beta_{b}$ should satisfy the solar system constraint
\cite{solarsystem}%
\begin{equation}
\beta_{b}\lesssim10^{-2}\text{ .} \label{ss}%
\end{equation}
We assume $\beta_{b}\equiv0$, which trivially satisfy the constraint (\ref{ss}).

We consider two kinds of scalar fields: the canonical scalar field, or
quintessence field, for which%
\begin{equation}
\mathcal{L}_{\varphi}\left(  x\right)  =\frac{1}{2}\partial_{\mu}%
\varphi\partial^{\mu}\varphi-V(\varphi)\text{ ,} \label{canonical}%
\end{equation}
and the tachyon scalar field, for which%
\begin{equation}
\mathcal{L}_{\varphi}\left(  x\right)  =-V(\varphi)\sqrt{1-\alpha\partial
^{\mu}\varphi\partial_{\mu}\varphi}\text{ ,} \label{tachyonic}%
\end{equation}
where $\alpha$ is a constant with dimension $MeV^{-4}$. Notice that in both
cases, we assume a Yukawa coupling with the dark matter field $\Psi$.

\subsection{Quintessence field}

For the quintessence field, $\mathcal{L}_{\varphi}\left(  x\right)  $ in the
action (\ref{action}) is given by (\ref{canonical}). From a variational
principle, we obtain%

\begin{equation}
i\gamma^{\mu}\nabla_{\mu}\Psi-M^{\ast}\Psi=0\text{ ,} \label{dirac}%
\end{equation}%
\begin{equation}
i(\nabla_{\mu}\bar{\Psi})\gamma^{\mu}+M^{\ast}\bar{\Psi}=0\text{ ,}
\label{diracadj}%
\end{equation}
where $M^{\ast}\equiv M-\beta\varphi$, and%
\begin{equation}
\nabla_{\mu}\partial^{\mu}\varphi+\frac{dV(\varphi)}{d\varphi}=\beta\bar{\Psi
}\Psi\text{ .} \label{eqmov_scalar}%
\end{equation}
Eqs. (\ref{dirac}) and (\ref{diracadj}) are, respectively, the covariant Dirac
equation and its adjoint, in the case of a non vanishing interaction between
the Dirac field and the scalar field $\varphi$. For homogeneous fields and
adopting the Friedmann-Robertson-Walker metric, $g_{\mu\nu}$=diag$\left(
1,-a^{2}\left(  t\right)  ,-a^{2}\left(  t\right)  ,-a^{2}\left(  t\right)
\right)  $, where $a^{2}\left(  t\right)  $ is the scale factor, eqs.
(\ref{dirac}) and (\ref{diracadj}) lead to%
\[
\frac{d(a^{3}\bar{\Psi}\Psi)}{dt}=0
\]
which is equivalent to%
\begin{equation}
\bar{\Psi}\Psi=\bar{\Psi}_{0}\Psi_{0}\left(  \frac{a_{0}}{a}\right)  ^{3}
\label{conser_psibarpsi}%
\end{equation}
and (\ref{eqmov_scalar}) reduces to%
\begin{equation}
\ddot{\varphi}+3H\dot{\varphi}+\frac{dV(\varphi)}{d\varphi}=\beta\bar{\Psi
}\Psi\text{ ,} \label{homoscalar}%
\end{equation}
where $H\equiv\frac{\dot{a}}{a}$ is the Hubble parameter.

From the energy-momentum tensor, we get%
\begin{align}
\rho_{\varphi}  &  =\frac{1}{2}\dot{\varphi}^{2}+V(\varphi)\text{
,}\label{rofiscalar}\\
P_{\varphi}  &  =\frac{1}{2}\dot{\varphi}^{2}-V(\varphi)\text{ ,}%
\label{pfiscalar}\\
\rho_{\Psi}  &  =M^{\ast}\bar{\Psi}\Psi\text{ ,}\label{ropsi}\\
P_{\Psi}  &  =0\text{ .}\nonumber
\end{align}
From (\ref{rofiscalar}) and (\ref{pfiscalar}) we have $\omega_{\varphi}%
\equiv\frac{P_{\varphi}}{\rho_{\varphi}}=\frac{\frac{1}{2}\dot{\varphi}%
^{2}-V(\varphi)}{\frac{1}{2}\dot{\varphi}^{2}+V(\varphi)}$. Deriving
(\ref{rofiscalar}) and (\ref{ropsi}) with respect to time and using
(\ref{conser_psibarpsi}) and (\ref{homoscalar}), we obtain%
\begin{equation}
\dot{\rho}_{\varphi}+3H\rho_{\varphi}(\omega_{\varphi}+1)=\beta\dot{\varphi
}\bar{\Psi}_{0}\Psi_{0}\left(  \frac{a_{0}}{a}\right)  ^{3}
\label{conser_rofi}%
\end{equation}
and%
\begin{equation}
\dot{\rho}_{\Psi}+3H\rho_{\Psi}=-\beta\dot{\varphi}\bar{\Psi}_{0}\Psi
_{0}\left(  \frac{a_{0}}{a}\right)  ^{3}\text{ ,} \label{conser_ropsi}%
\end{equation}
where the dot represents derivative with respect to time.

For baryonic matter and radiation, we have respectively%
\begin{equation}
\dot{\rho}_{b}+3H\rho_{b}=0 \label{conserbaryon}%
\end{equation}
and%
\begin{equation}
\dot{\rho}_{r}+3H\rho_{r}(\omega_{r}+1)=0\text{ ,} \label{conserrad}%
\end{equation}
where $\omega_{r}=\frac{1}{3}$. Eqs. (\ref{conserbaryon}) and (\ref{conserrad}%
) implies that $\rho_{b}=\frac{\rho_{b0}}{a^{3}}$ and $\rho_{r}=\frac
{\rho_{r0}}{a^{4}}$, respectively. The subscript $0$ denotes the quantities
today. We are considering the radiation as composed by photons and massless
neutrinos, so that $\rho_{r0}=\left(  1+0.2271N_{eff}\right)  \rho_{\gamma0}$,
where $N_{eff}=3.04$ is the effective number of relativistic degrees of
freedom and $\rho_{\gamma0}$ is the energy density of photons, given
by\textbf{ }$\rho_{\gamma0}=\frac{\pi^{2}}{15}T_{CMB}$, being $T_{CMB}=2.725K$
the CMB temperature today. The Friedmann equation for a flat universe reads%
\begin{equation}
H^{2}=\frac{1}{3M_{Pl}^{2}}\left[  M^{\ast}\bar{\Psi}_{0}\Psi_{0}\left(
\frac{a_{0}}{a}\right)  ^{3}+\frac{1}{2}\dot{\varphi}^{2}+V(\varphi
)+\frac{\rho_{b0}}{a^{3}}+\frac{\rho_{r0}}{a^{4}}\right]  \text{ .}
\label{friedmann_sc}%
\end{equation}

In order to determine the dynamics of the interacting quintessence field, it
is necessary to specify the potential $V(\varphi)$. Instead of choosing an
explicit form for $V(\varphi)$, we will specify it implicitly, by imposing
that the energy density of the quintessence field, given by (\ref{rofiscalar}%
), must match the holographic dark energy density, $\rho_{\Lambda}%
=3c^{2}M_{Pl}^{2}L^{-2}$, where $c$ is a numerical constant and $L$ is the
infrared cutoff. The evolution of the interacting quintessence field with
redshift will be given by the equation of evolution for the holographic dark
energy density, with a certain expression for the equation of state parameter
$\omega_{\varphi}$. In fact, we will see that imposing the energy density of
the quintessence field to match the holographic dark energy density leads to
an expression for the potential.

In \cite{Li} it has been argued that, in order that holographic dark energy
drives the recent period of accelerated expansion, the IR cutoff $L$ must be
the event horizon $R_{h}=a\left(  t\right)  \int_{t}^{\infty}\frac{dt^{\prime
}}{a\left(  t^{\prime}\right)  }$. Substituting $R_{h}$ in the expression of
the holographic dark energy, we get $R_{h}=\frac{c}{H\sqrt{\Omega_{\varphi}}}%
$, therefore,%
\[
\int_{t}^{\infty}\frac{dt^{\prime}}{a\left(  t^{\prime}\right)  }=\frac
{c}{a\left(  t\right)  H\sqrt{\Omega_{\varphi}}}\text{ .}%
\]
Differentiating both sides with respect to time, using the Friedmann equation
(\ref{friedmann_sc}) together with conservation equations (\ref{conser_rofi})
and (\ref{conser_ropsi}), we obtain%
\begin{equation}
\frac{d\Omega_{\varphi}}{dz}=-\frac{\Omega_{\varphi}}{1+z}\left(  2\frac
{\sqrt{\Omega_{\varphi}}}{c}+3\Omega_{\varphi}\omega_{\varphi}+\Omega
_{r}+1\right)  \text{ .} \label{eq_mov_holo}%
\end{equation}
Equation (\ref{eq_mov_holo}) is just the equation of evolution for the
holographic dark energy \cite{Li}. Using the Friedmann equation
(\ref{friedmann_sc}), the conservation equations (\ref{conserbaryon}) and
(\ref{conserrad}) can be written as%
\begin{equation}
\frac{d\Omega_{b}}{dz}=-\frac{\Omega_{b}}{1+z}\left(  3\Omega_{\varphi}%
\omega_{\varphi}+\Omega_{r}\right)  \label{eq_movbar}%
\end{equation}
and%
\begin{equation}
\frac{d\Omega_{r}}{dz}=-\frac{\Omega_{r}}{1+z}\left(  3\Omega_{\varphi}%
\omega_{\varphi}+\Omega_{r}-1\right)  \text{ .} \label{eq_movrad}%
\end{equation}

We define $r\equiv\frac{\rho_{\Psi}}{\rho_{\varphi}}$. Deriving $r$ with
respect to time, using (\ref{conser_rofi}), (\ref{conser_ropsi}),
(\ref{rofiscalar}) and (\ref{pfiscalar}) we obtain%
\begin{equation}
\dot{r}=3Hr\omega_{\varphi}-sign\left[  \dot{\varphi}\right]  \frac
{\beta\left(  1+r\right)  }{\sqrt{3}M_{Pl}H}\sqrt{\frac{1+\omega_{\varphi}%
}{\Omega_{\varphi}}}\bar{\Psi}_{0}\Psi_{0}\left(  \frac{1+z}{1+z_{0}}\right)
^{3}\text{ .} \label{rdotscalar}%
\end{equation}

We can rewrite $\bar{\Psi}_{0}\Psi_{0}$ in terms of observable quantities. In
fact, by imposing that the dark matter density today matches the observed
value, we obtain $\bar{\Psi}_{0}\Psi_{0}=\frac{3M_{Pl}^{2}H_{0}^{2}\left(
1-\Omega_{\varphi0}-\Omega_{b0}-\Omega_{r0}\right)  }{M-\beta\phi_{0}}$. The
sign of $\dot{\varphi}$ is arbitrary, as it can be modified by redefinitions
of the field, $\varphi\rightarrow-\varphi$, and of the coupling constant,
$\beta\rightarrow-\beta$. Noticing that $r=\frac{1-\Omega_{\varphi}-\Omega
_{b}-\Omega_{r}}{\Omega_{\varphi}}$, we can substitute $r$ and $\dot{r}$ in
(\ref{rdotscalar}) by $\Omega_{\varphi}$, $\Omega_{b}$, $\Omega_{r}$,
$\dot{\Omega}_{\varphi}$, $\dot{\Omega}_{b}$ and $\dot{\Omega}_{r}$. Using
(\ref{eq_mov_holo}), (\ref{eq_movbar}) and (\ref{eq_movrad}) we obtain, after
some algebra%
\begin{equation}
\omega_{\varphi}\left(  z\right)  =-\frac{1}{3}-\frac{2\sqrt{\Omega_{\varphi}%
}}{3c}+\frac{\gamma\left(  z\right)  }{3}\left[  \gamma\left(  z\right)
+\sqrt{\gamma\left(  z\right)  ^{2}+4\left(  1-\frac{\sqrt{\Omega_{\varphi}}%
}{c}\right)  }\right]  \text{ ,} \label{wfi_final_sc}%
\end{equation}
where%
\begin{equation}
\gamma\left(  z\right)  =\frac{1}{\sqrt{2}}\delta M_{Pl}\frac{\left(
1-\Omega_{\varphi0}-\Omega_{b0}-\Omega_{r0}\right)  }{E\left(  z\right)
^{2}\sqrt{\Omega_{\varphi}}}\left(  \frac{1+z}{1+z_{0}}\right)  ^{3}\text{ ,}
\label{gama_sc}%
\end{equation}
with%
\begin{equation}
E\left(  z\right)  \equiv\frac{H\left(  z\right)  }{H_{0}}=\sqrt{\frac{\left[
\left(  1-\delta\Delta\varphi\right)  \left(  1-\Omega_{\varphi0}-\Omega
_{b0}-\Omega_{r0}\right)  +\Omega_{b0}\right]  }{1-\Omega_{\varphi}}\left(
\frac{1+z}{1+z_{0}}\right)  ^{3}+\frac{\Omega_{r0}}{1-\Omega_{\varphi}}\left(
\frac{1+z}{1+z_{0}}\right)  ^{4}}\text{ ,} \label{E_final_sc}%
\end{equation}
where $\Delta\varphi\left(  z\right)  \equiv\varphi\left(  z\right)
-\varphi_{0}$ and $\delta\equiv\frac{\beta}{M-\beta\varphi_{0}}$ is an
effective coupling constant. Notice that, if $\delta=0$, (\ref{wfi_final_sc})
reproduces the equation of state parameter obtained in \cite{Li}.

The evolution of the quintessence field is given by%
\begin{equation}
\frac{d\varphi}{dz}=-\frac{\sqrt{3}M_{Pl}\sqrt{\Omega_{\varphi}\left(
z\right)  \left(  1+\omega_{\varphi}\left(  z\right)  \right)  }}{1+z}
\label{eq_fi_sc}%
\end{equation}
From (\ref{eq_mov_holo}), (\ref{eq_movbar}), (\ref{eq_movrad}) and
(\ref{eq_fi_sc}) we can calculate the evolution with redshift of all
observables in the model. If we wish to calculate the time dependence, we need
to integrate the Friedmann equation (\ref{friedmann_sc}), which can be written
in the form%
\[
\frac{dt}{dz}=-\frac{1}{H_{0}E(z)(1+z)}\text{ .}%
\]
From (\ref{rofiscalar}), we can compute the potential $V\left(  z\right)  $ as%
\begin{equation}
\frac{V\left(  z\right)  }{\rho_{c0}}=\frac{E^{2}(z)\Omega_{\varphi}(z)}%
{2}\left(  1-\omega_{\varphi}(z)\right)  \text{ ,} \label{potential_sc}%
\end{equation}
where $\rho_{c0}=3M_{Pl}^{2}H_{0}^{2}$, $E\left(  z\right)  $ is given by
(\ref{E_final_sc}), $\omega_{\varphi}\left(  z\right)  $ is given by
(\ref{wfi_final_sc}) and $\Omega_{\varphi}\left(  z\right)  $ is the solution
of (\ref{eq_mov_holo}). From (\ref{potential_sc}) and (\ref{eq_fi_sc}), we can
compute $V(\varphi)$.

Here it is worth saying that in the holographic dark energy model, in the non
interacting case - (\ref{wfi_final_sc}) with $\delta=0$ - $\omega_{\varphi}$
can be less than $-1$. However, as already mentioned in \cite{escalarholo}, if
we wish that the holographic dark energy is the quintessence field, then
because (\ref{eq_fi_sc}), $\omega_{\varphi}$ must be more than $-1$.
Nevertheless, in the interacting case considered here, due to the fact that
$\omega_{\varphi}$ depends explicitly on $\varphi$, $\omega_{\varphi}$ can not
be less than $-1$. On the other hand, the square root in (\ref{wfi_final_sc})
must be real. We can verify that $\omega_{\varphi}$ is real and $\omega
_{\varphi}>-1$ if (i) $\frac{\sqrt{\Omega_{\varphi0}}}{c}<1$ or (ii)
$\frac{\sqrt{\Omega_{\varphi0}}}{c}>1$ and $\left\vert \delta\right\vert
M_{Pl}\geq\frac{2\sqrt{2\Omega_{\varphi0}}}{1-\Omega_{\varphi0}-\Omega
_{b0}-\Omega_{r0}}\sqrt{\frac{\sqrt{\Omega_{\varphi0}}}{c}-1}$. However, case
(ii) is irrelevant, as it corresponds to large values of $\left\vert
\delta\right\vert M_{Pl}$. For example, if $\Omega_{\varphi0}=0.7$ and
$c=0.8$, we have $\left\vert \delta\right\vert M_{Pl}\gtrsim1.69$. Below, we
will see that the observational data constrain $\left\vert \delta\right\vert
M_{Pl}\sim10^{-1}$. In order that $\omega_{\varphi}$ be real for all future
times, as $\Omega_{\varphi}\rightarrow1$, it is necessary that $c\geq1$.

It is interesting to notice that the condition $\frac{\sqrt{\Omega_{\varphi0}%
}}{c}<1$ is precisely the same one for which the entropy of the universe
increases \cite{Li}. As $\Omega_{\varphi}\rightarrow1$ in the future, it is
necessary that $c\geq1$. Therefore, the condition for $\omega_{\varphi}$ be
real is precisely the same one for the entropy to increase. So the model
respects the second law of thermodynamics.

\begin{figure}[ptb]
\begin{center}
\includegraphics[width=6.23cm,height=5.0cm]{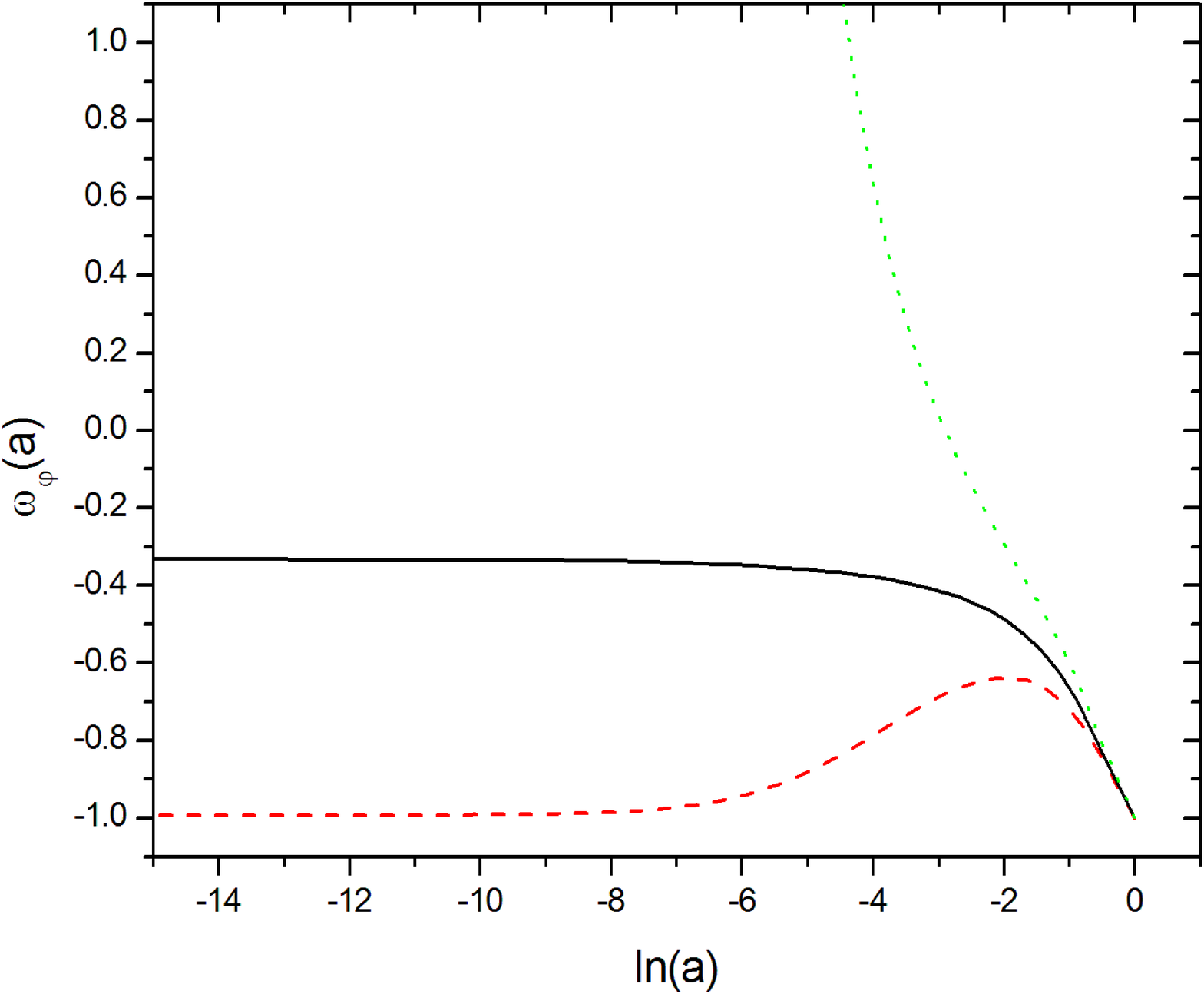}
\ \includegraphics[width=6.23cm,height=5.0cm]{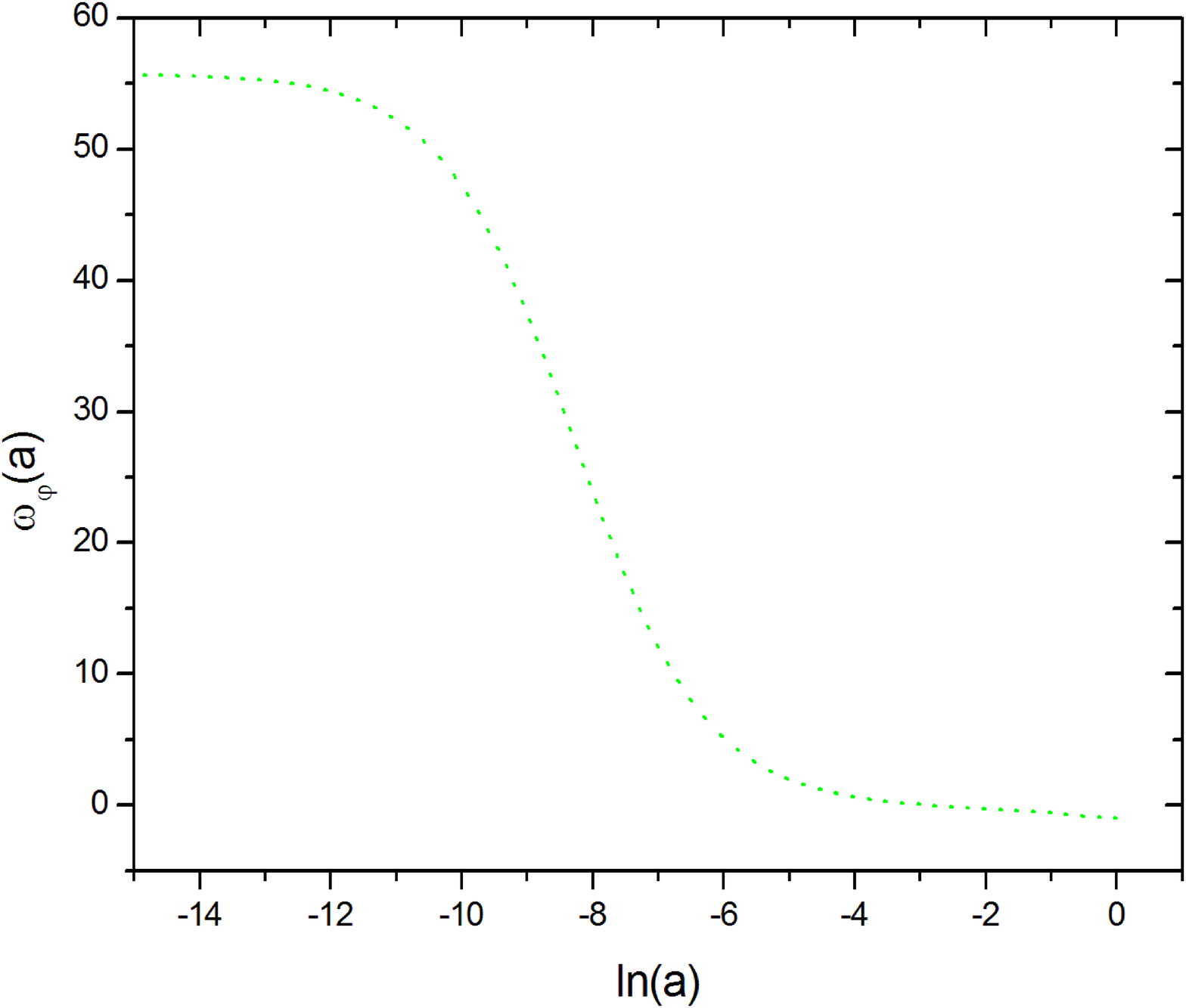}
\end{center}
\caption{(a) Equation of state parameter of the holographic quintessence
model, for $c=0.85$ and $\delta M_{Pl}=-0.1$ (red dashed line), $\delta
M_{Pl}=0$ (black solid line) and $\delta M_{Pl}=+0.1$ (green dotted line). (b)
Full range of the equation of state parameter for $\delta M_{Pl}=+0.1$.}%
\label{fig_w_escrad}%
\end{figure}

In figure \ref{fig_w_escrad} we see the evolution of the equation of state
parameter $\omega_{\varphi}$ with the scale factor $a$. For the non
interacting case, $\delta=0$, we have $\omega_{\varphi}\rightarrow-1/3$, as
$\Omega_{\varphi}(z)\ll1$ for $z\gg1$. For $\delta<0$, $\omega_{\varphi}>-1$
in the matter era, then approaches $-1$ in the radiation era. For $\delta>0$,
$\omega_{\varphi}$ will eventually turns out positive and possibly
$\omega_{\varphi}\gg1$ in very early times, as in the case showed in figure
\ref{fig_w_escrad}. This behaviour is explained as follows. In the matter era,
$E^{2}(z)\sim\left(  1+z\right)  ^{3}$ so that $\gamma\left(  z\right)
\sim\frac{1}{\sqrt{\Omega_{\varphi}}}$. From (\ref{eq_mov_holo})
$\frac{d\Omega_{\varphi}}{dz}<0$, so $\left\vert \gamma\left(  z\right)
\right\vert $ increases with redshift. This increasing of $\left\vert
\gamma\left(  z\right)  \right\vert $ will continue until the radiation era,
when $E^{2}(z)\sim\left(  1+z\right)  ^{4}$ and $\Omega_{\varphi}%
(z)\sim\left(  1+z\right)  ^{-2}$ so that $\left\vert \gamma\left(  z\right)
\right\vert \rightarrow cte$. Tipically this constant will be much more than
one. Therefore, for high redshifts $\omega_{\varphi}\left(  z\right)
\simeq-\frac{1}{3}+\frac{\gamma}{3}\left[  \gamma+\left\vert \gamma\right\vert
+\frac{2}{\left\vert \gamma\right\vert }\right]  $. If $\delta<0$ then
$\gamma<0$ and $\omega_{\varphi}\left(  z\right)  \simeq-1$ in the radiation
era. If $\delta>0$ then $\gamma>0$ and $\omega_{\varphi}\left(  z\right)
\simeq\frac{1}{3}+\frac{2}{3}\gamma^{2}\rightarrow cte$. Notice from
(\ref{potential_sc}) that if $\omega_{\varphi}\left(  z\right)  >1$ then
$V<0$. In order to avoid it, we impose the condition $\omega_{\varphi}\leq1$
for all $z$. This condition is satisfied if $\delta M_{Pl}\lesssim\frac
{\sqrt{2}\left(  1-\Omega_{\varphi0}\right)  }{1-\Omega_{\varphi0}-\Omega
_{b0}}\sqrt{\Omega_{\varphi}}$. As we have $\Omega_{\varphi}\ll1$ as $z$
increase, there will be an abrupt decrease in the positive tail of the
probability distribuction of $\delta$, as well as in the confidence regions of
$\delta$ with other parameters.

\begin{figure}[ptb]
\begin{center}
\includegraphics[width=6.23cm,height=5.0cm]{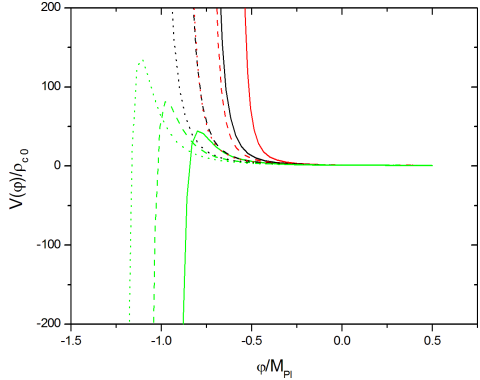}
\end{center}
\caption{Potential of the holographic quintessence field $V(\varphi)$, in
units of $\rho_{c0}=3M_{Pl}^{2}H_{0}^{2}$. $\varphi$ is in units of $M_{Pl}$.
The solid lines are for $c=0.95$, the dashed ones are for $c=1.1$ and the
dotted are for $c=1.25$. For each value of $c$ the curves from right to left
are for $\delta M_{Pl}=-0.1$ (red), $\delta M_{Pl}=0$ (black) and $\delta
M_{Pl}=+0.1$ (green), respectively.}%
\label{potential_canonico}%
\end{figure}

In figure \ref{potential_canonico}\textbf{,} $V(\varphi)$ is shown for some
values of $\delta$ and $c$. Notice that there is a region where $V(\varphi)$
is almost constant, that is, there is a slow-roll region. As we chose
$\dot{\varphi}$ positive, then $\varphi$ evolves to this slow-roll region.
However, if we had chosen $\dot{\varphi}$ negative, then because the right
hand side of (\ref{eq_fi_sc}) would have the opposite sign, so $\frac
{dV(\varphi)}{d\varphi}$ would have also the opposite sign and again $\varphi$
would evolve to the slow-roll region. Notice also that for $\delta
M_{Pl}=+0.1$, the potential is negative in the past.

The equation for evolution of $\varphi$ (\ref{eq_fi_sc}) can be written in an
integral form as%
\[
\Delta\varphi\left(  z\right)  =-\sqrt{3}M_{Pl}\int_{0}^{z}\frac{\sqrt
{\Omega_{\varphi}(z)\left(  1+\omega_{\varphi}\left(  z\right)  \right)  }%
}{1+z}dz\text{ .}%
\]
Since the model depends on $\Delta\varphi$ - through $E\left(  z\right)  $ -
and neither on $\varphi$ nor on $\varphi_{0}$, then it is independent of
$\varphi_{0}$. In other words, $\varphi_{0}$ is not a parameter of the model
and can be chosen arbitrarily. Therefore, the parameters of the model are
$\delta$, $c$, $h,$ $\Omega_{b0}$ and $\Omega_{\varphi0}$.

\subsection{Tachyon scalar field}

In the case of dark energy modeled as the tachyon scalar field, $\mathcal{L}%
_{\varphi}\left(  x\right)  $ in the action (\ref{action}) is given by
(\ref{tachyonic}). From a variational principle, we obtain%
\begin{equation}
i\gamma^{\mu}\nabla_{\mu}\Psi-M^{\ast}\Psi=0\text{ ,} \label{dirac_tac}%
\end{equation}%
\begin{equation}
i(\nabla_{\mu}\bar{\Psi})\gamma^{\mu}+M^{\ast}\bar{\Psi}=0\text{ ,}
\label{diracadj_tac}%
\end{equation}
where $M^{\ast}\equiv M-\beta\varphi$, and%
\begin{equation}
\nabla_{\mu}\partial^{\mu}\varphi+\alpha\frac{\partial^{\mu}\varphi
(\nabla_{\mu}\partial_{\sigma}\varphi)\partial^{\sigma}\varphi}{1-\alpha
\partial_{\mu}\varphi\partial^{\mu}\varphi}+\frac{1}{\alpha}\frac
{dlnV(\varphi)}{d\varphi}=\frac{\beta\bar{\Psi}\Psi}{\alpha V(\varphi)}%
\sqrt{1-\alpha\partial^{\mu}\varphi\partial_{\mu}\varphi}\text{ .}
\label{eqmov_taquions}%
\end{equation}
The equations of motion for $\Psi$ and $\bar{\Psi}$ (\ref{dirac_tac}) and
(\ref{diracadj_tac}) are the interacting covariant Dirac equation and its
adjoint, respectively, i. e., (\ref{dirac_tac}) and (\ref{diracadj_tac}) are
almost the same as eqs. (\ref{dirac}) and (\ref{diracadj}), the only
difference is that the scalar field $\varphi$ in $M^{\ast}$ now is the tachyon
field. For homogeneous fields and adopting the Friedmann-Robertson-Walker
metric, $g_{\mu\nu}$=diag$\left(  1,-a^{2}\left(  t\right)  ,-a^{2}\left(
t\right)  ,-a^{2}\left(  t\right)  \right)  $, where $a^{2}\left(  t\right)  $
is the scale factor, (\ref{eqmov_taquions}) reduces to%
\begin{equation}
\ddot{\varphi}=-(1-\alpha\dot{\varphi}^{2})\left[  \frac{1}{\alpha}%
\frac{dlnV(\varphi)}{d\varphi}+3H\dot{\varphi}-\frac{\beta\bar{\Psi}\Psi
}{\alpha V(\varphi)}\sqrt{1-\alpha\dot{\varphi}^{2}}\right]  \text{ ,}
\label{homotaq}%
\end{equation}
whereas for the fermions, the equations of motion will reduce to eq.
(\ref{conser_psibarpsi}), as already obtained above:%
\begin{equation}
\bar{\Psi}\Psi=\bar{\Psi}_{0}\Psi_{0}\left(  \frac{a_{0}}{a}\right)
^{3}\text{ .} \tag{8}%
\end{equation}

From the energy-momentum tensor, we get%
\begin{align}
\rho_{\varphi}  &  =\frac{V(\varphi)}{\sqrt{1-\alpha\dot{\varphi}^{2}}}\text{
,}\label{rofi}\\
P_{\varphi}  &  =-V(\varphi)\sqrt{1-\alpha\dot{\varphi}^{2}}\text{
,}\label{pfi}\\
\rho_{\Psi}  &  =M^{\ast}\bar{\Psi}\Psi\text{ ,}\nonumber\\
P_{\Psi}  &  =0\text{ .}\nonumber
\end{align}
From (\ref{rofi}) and (\ref{pfi}) we have $\omega_{\varphi}\equiv
\frac{P_{\varphi}}{\rho_{\varphi}}=\alpha\dot{\varphi}^{2}-1$. Deriving
(\ref{rofi}) and (\ref{pfi}) with respect to time and using (\ref{homotaq})
and (\ref{conser_psibarpsi}), we get%

\begin{equation}
\dot{\rho}_{\varphi}+3H\rho_{\varphi}(\omega_{\varphi}+1)=\beta\dot{\varphi
}\bar{\Psi}_{0}\Psi_{0}\left(  \frac{a_{0}}{a}\right)  ^{3}
\label{conser_rotac}%
\end{equation}
and%
\begin{equation}
\dot{\rho}_{\Psi}+3H\rho_{\Psi}=-\beta\dot{\varphi}\bar{\Psi}_{0}\Psi
_{0}\left(  \frac{a_{0}}{a}\right)  ^{3}\text{ ,} \label{conser_psitac}%
\end{equation}
where the dot represents derivative with respect to time.

For baryonic matter and radiation, the conservation equations are the same as
in the quintessence model. We have
\begin{equation}
\dot{\rho}_{b}+3H\rho_{b}=0 \tag{15}%
\end{equation}
and%
\begin{equation}
\dot{\rho}_{r}+3H\rho_{r}(\omega_{r}+1)=0\text{ ,} \tag{16}%
\end{equation}
where $\omega_{r}=\frac{1}{3}$. Eqs. (\ref{conserbaryon}) and (\ref{conserrad}%
) implies that $\rho_{b}=\frac{\rho_{b0}}{a^{3}}$ and $\rho_{r}=\frac
{\rho_{r0}}{a^{4}}$, respectively. The subscript $0$ denotes the quantities
today. We are considering the radiation as composed by photons and massless
neutrinos, so that $\rho_{r0}=\left(  1+0.2271N_{eff}\right)  \rho_{\gamma0}$,
where $N_{eff}=3.04$ is the effective number of relativistic degrees of
freedom and $\rho_{\gamma0}$ is the energy density of photons, given by
$\rho_{\gamma0}=\frac{\pi^{2}}{15}T_{CMB}$, being $T_{CMB}=2.725K$ the CMB
temperature today. The Friedmann equation for a flat universe reads%

\begin{equation}
H^{2}=\frac{1}{3M_{Pl}^{2}}\left[  M^{\ast}\bar{\Psi}_{0}\Psi_{0}\left(
\frac{a_{0}}{a}\right)  ^{3}+\frac{V(\varphi)}{\sqrt{1-\alpha\dot{\varphi}%
^{2}}}+\frac{\rho_{b0}}{a^{3}}+\frac{\rho_{r0}}{a^{4}}\right]  \text{ .}
\label{friedmann_tac}%
\end{equation}

Now, as done in the case of the quintessence field, we will identify the
tachyon energy density (\ref{rofi}) with the holographic dark energy density
$\rho_{\Lambda}=3c^{2}M_{Pl}^{2}L^{-2}$. From a similar reasoning, we obtain
again the equations (\ref{eq_mov_holo}), (\ref{eq_movbar}) and
(\ref{eq_movrad}),%
\begin{equation}
\frac{d\Omega_{\varphi}}{dz}=-\frac{\Omega_{\varphi}}{1+z}\left(  2\frac
{\sqrt{\Omega_{\varphi}}}{c}+3\Omega_{\varphi}\omega_{\varphi}+\Omega
_{r}+1\right)  \text{ ,} \tag{18}%
\end{equation}%
\begin{equation}
\frac{d\Omega_{b}}{dz}=-\frac{\Omega_{b}}{1+z}\left(  3\Omega_{\varphi}%
\omega_{\varphi}+\Omega_{r}\right)  \tag{19}%
\end{equation}
and%
\begin{equation}
\frac{d\Omega_{r}}{dz}=-\frac{\Omega_{r}}{1+z}\left(  3\Omega_{\varphi}%
\omega_{\varphi}+\Omega_{r}-1\right)  \text{ .} \tag{20}%
\end{equation}
Also we obtain%
\begin{equation}
\dot{r}=3Hr\omega_{\varphi}-sign\left[  \dot{\varphi}\right]  \frac
{\beta\left(  1+r\right)  ^{2}\sqrt{1+\omega_{\varphi}}}{3M_{Pl}^{2}%
\sqrt{\alpha}H^{2}}\bar{\Psi}_{0}\Psi_{0}\left(  \frac{1+z}{1+z_{0}}\right)
^{3}\text{ ,} \label{rdottac}%
\end{equation}
with $r\equiv\frac{\rho_{\Psi}}{\rho_{\varphi}}$. The sign of $\dot{\varphi}$
is arbitrary, as it can be modified by redefinitions of the field,
$\varphi\rightarrow-\varphi$, and of the coupling constant, $\beta
\rightarrow-\beta$. We can rewrite $\bar{\Psi}_{0}\Psi_{0}$ in terms of
observable quantities, by imposing that the dark matter density today matches
the observed value. We obtain $M\bar{\Psi}_{0}\Psi_{0}=\frac{3M_{Pl}^{2}%
H_{0}^{2}\left(  1-\Omega_{\phi0}-\Omega_{b0}-\Omega_{r0}\right)  }%
{1-\frac{\beta}{M\sqrt{\alpha}}\phi_{0}}$, where we defined $\phi\equiv
\sqrt{\alpha}\varphi$. Furthermore, noticing that $r=\frac{1-\Omega_{\varphi
}-\Omega_{b}-\Omega_{r}}{\Omega_{\varphi}}$, we can eliminate $r$ and $\dot
{r}$ in favor of $\Omega_{\varphi}$, $\Omega_{b}$, $\Omega_{r}$, $\dot{\Omega
}_{\varphi}$, $\dot{\Omega}_{b}$ and $\dot{\Omega}_{r}$ in (\ref{rdottac}).
Using (\ref{eq_mov_holo}), (\ref{eq_movbar}) and (\ref{eq_movrad}) we obtain,
after some algebra%
\begin{equation}
\omega_{\phi}\left(  z\right)  =-\frac{1}{3}-\frac{2\sqrt{\Omega_{\phi}\left(
z\right)  }}{3c}+\frac{\gamma\left(  z\right)  }{3}\left[  \gamma\left(
z\right)  +\sqrt{\gamma\left(  z\right)  ^{2}+4\left(  1-\frac{\sqrt
{\Omega_{\phi}\left(  z\right)  }}{c}\right)  }\right]  \text{ ,}
\label{wfi_final_tac}%
\end{equation}
where%
\begin{equation}
\gamma\left(  z\right)  \equiv\frac{1}{\sqrt{6}}\frac{\delta}{H_{0}}%
\frac{1-\Omega_{\phi0}-\Omega_{b0}-\Omega_{r0}}{\Omega_{\phi}\left(  z\right)
E^{3}\left(  z\right)  }\left(  \frac{1+z}{1+z_{0}}\right)  ^{3}\text{ ,}
\label{gama_tac}%
\end{equation}
with%
\begin{equation}
E\left(  z\right)  \equiv\frac{H\left(  z\right)  }{H_{0}}=\sqrt{\frac{\left[
\left(  1-\delta\Delta\phi\right)  \left(  1-\Omega_{\phi0}-\Omega_{b0}%
-\Omega_{r0}\right)  +\Omega_{b0}\right]  }{1-\Omega_{\phi}}\left(  \frac
{1+z}{1+z_{0}}\right)  ^{3}+\frac{\Omega_{r0}}{1-\Omega_{\phi}}\left(
\frac{1+z}{1+z_{0}}\right)  ^{4}}\text{ ,} \label{E_final_tac}%
\end{equation}
where $\Delta\phi\left(  z\right)  \equiv\phi\left(  z\right)  -\phi_{0}$ and
$\delta\equiv\frac{\frac{\beta}{M\sqrt{\alpha}}}{1-\frac{\beta}{M\sqrt{\alpha
}}\phi_{0}}$ is an effective coupling constant. As in the quintessence field
case, if $\delta=0$, (\ref{wfi_final_tac}) reproduces the equation of state
parameter obtained in \cite{Li}.

The evolution of the tachyon scalar field is given by%
\begin{equation}
\frac{d\phi}{dz}=-\frac{\sqrt{1+\omega_{\phi}\left(  z\right)  }}%
{H_{0}E\left(  z\right)  \left(  1+z\right)  }\text{ .} \label{eq_fi_tac}%
\end{equation}
From (\ref{eq_mov_holo}), (\ref{eq_movbar}), (\ref{eq_movrad}) and
(\ref{eq_fi_tac}) we can calculate the evolution with redshift of all
observables in the model. If we wish to calculate the time dependence, we need
to integrate the Friedmann equation (\ref{friedmann_tac}), which can be
written in the form%
\[
\frac{dt}{dz}=-\frac{1}{H_{0}E(z)(1+z)}\text{ .}%
\]
From (\ref{rofi}), we can compute the potential $V\left(  z\right)  $ as%
\begin{equation}
\frac{V\left(  z\right)  }{\rho_{c0}}=E^{2}\left(  z\right)  \Omega_{\phi
}\left(  z\right)  \sqrt{-\omega_{\phi}\left(  z\right)  }\text{ ,}
\label{potential}%
\end{equation}
where $\rho_{c0}=3M_{Pl}^{2}H_{0}^{2}$, $E\left(  z\right)  $ is given by
(\ref{E_final_tac}), $\omega_{\phi}\left(  z\right)  $ is given by
(\ref{wfi_final_tac}) and $\Omega_{\phi}\left(  z\right)  $ is the solution of
(\ref{eq_mov_holo}). From (\ref{potential}) and (\ref{eq_fi_tac}), we can
compute $V(\phi)$.

The square root in (\ref{wfi_final_tac}) must be real. Furthermore, in
analogous manner to the quintessence model, $\omega_{\phi}$ must be more than
$-1$ because (\ref{eq_fi_tac}). We can verify that $\omega_{\phi}$ is real and
$\omega_{\phi}>-1$ if (i) $\frac{\sqrt{\Omega_{\phi0}}}{c}<1$ or (ii)
$\frac{\sqrt{\Omega_{\phi0}}}{c}>1$ and $\frac{\left\vert \delta\right\vert
}{H_{0}}>2\sqrt{6}\frac{\Omega_{\phi0}}{1-\Omega_{\phi0}-\Omega_{b0}%
-\Omega_{r0}}\sqrt{\frac{\sqrt{\Omega_{\phi0}}}{c}-1}$. However, case (ii) is
irrelevant, as it corresponds to large values of $\frac{\left\vert
\delta\right\vert }{H_{0}}$. For example, if $\Omega_{\phi0}=0.7$ and $c=0.8$,
we have $\frac{\left\vert \delta\right\vert }{H_{0}}\gtrsim2.45$. Below, we
will see that the observational data constrain $\frac{\left\vert
\delta\right\vert }{H_{0}}\sim10^{-1}$. In order that $\omega_{\phi}$ be real
for all future times, as $\Omega_{\phi}\rightarrow1$, it is necessary that
$c\geq1$. As already mentioned in the quintessence case, this is also the
condition for the entropy to increase for all future times, so the tachyon
model also respects the second law of thermodynamics.

\begin{figure}[ptb]
\begin{center}
\includegraphics[width=6.23cm,height=5.0cm]{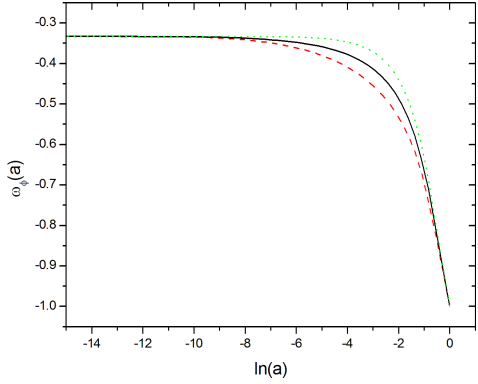}
\end{center}
\caption{Equation of state parameter of the holographic tachyon model, for
$c=0.85$ and $\frac{\delta}{H_{0}}=-0.1$ (red dashed line), $\frac{\delta
}{H_{0}}=0$ (black solid line) and $\frac{\delta}{H_{0}}=+0.1$ (green dotted
line).}%
\label{fig_w_taqrad}%
\end{figure}

The evolution of the equation of state parameter $\omega_{\phi}$ is showed in
figure \ref{fig_w_taqrad}. We have $\omega_{\phi}\rightarrow-1/3$ for $z\gg1$.
In the non interacting case - $\delta=0$ - this occurs simply because
$\Omega_{\phi}(z)\ll1$ in high redshifts. The behaviour for the interacting
case - $\delta\neq0$ - is explained as follows. In the matter era
$E^{2}(z)\sim\left(  1+z\right)  ^{3}$ so that $\gamma\left(  z\right)
\sim\frac{1}{\Omega_{\phi}\left(  z\right)  \left(  1+z\right)  ^{1,5}}$.
Using (\ref{eq_mov_holo}) we infer $\frac{d\left\vert \gamma\left(  z\right)
\right\vert }{dz}>0$, that is $\left\vert \gamma\left(  z\right)  \right\vert
$ increases with redshift $z$. Therefore, if $\delta<0$ then $\gamma\left(
z\right)  <0$ and $\omega_{\phi}$ increases slower than in the non interacting
case. If $\delta>0$ then $\gamma\left(  z\right)  >0$ and $\omega_{\phi}$
increases faster than in the non interacting case. In the radiation era
$\gamma\left(  z\right)  \sim\frac{1}{1+z}$ and it turns out to be negligible,
so that $\omega_{\phi}\simeq-1/3$.

\begin{figure}[ptb]
\begin{center}
\includegraphics[width=6.23cm,height=5.0cm]{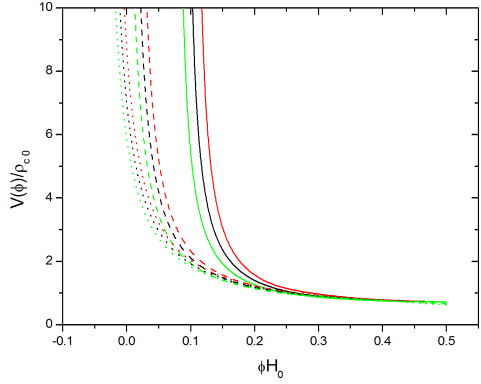}
\end{center}
\caption{Potential of the holographic tachyon field $V(\phi)$, in units of
$\rho_{c0}=3M_{Pl}^{2}H_{0}^{2}$. $\phi$ is in units of $H_{0}^{-1}$. The
solid lines are for $c=0.85$, the dashed ones are for $c=1.1$ and the dotted
are for $c=1.35$. For each value of $c$ the curves from right to left are for
$\frac{\delta}{H_{0}}=-0.1$ (red), $\frac{\delta}{H_{0}}=0$ (black) and
$\frac{\delta}{H_{0}}=+0.1$ (green), respectively.}%
\label{potential_taq}%
\end{figure}

In figure \ref{potential_taq}, $V(\phi)$ is shown for some values of $\delta$
and $c$. Notice that - as in the case of the quintessence field potential -
there is a region where $V(\phi)$ is almost constant, that is, there is a
slow-roll region. As we chose $\dot{\phi}$ positive, then $\phi$ evolves to
this slow-roll region. However, if we had chosen $\dot{\phi}$ negative, then
because the right hand side of (\ref{eq_fi_tac}) would have the opposite sign,
so $\frac{dV(\phi)}{d\phi}$ would have also the opposite sign and again $\phi$
would evolve to the slow-roll region.

The equation for evolution of $\phi$ (\ref{eq_fi_tac}) can be written in an
integral form as%
\[
\Delta\phi=-\frac{1}{H_{0}}\int_{0}^{z}\frac{\sqrt{1+\omega_{\phi}\left(
z\right)  }}{E(z)(1+z)}dz\text{ .}%
\]

As before, the model depends on $\Delta\phi$ - through $E\left(  z\right)  $ -
and neither on $\phi$ nor on $\phi_{0}$, then it is independent of $\phi_{0}$.
Therefore, the parameters of the model are $\delta$, $c$, $h,$ $\Omega_{b0}$
and $\Omega_{\phi0}$. Below, we discuss the comparison with observational data
and the results obtained.

\section{Constraints from observational data}

In \cite{lookback}, the lookback time method has been discussed. Given an
object $i$ at redshift $z_{i}$, its age $t(z_{i})$ is defined as the
difference between the age of the universe at $z_{i}$ and the age of the
universe at the formation redshift of the object, $z_{F}$, that is,
\begin{align}
t(z_{i})  &  =H_{0}^{-1}\left[  \int_{z_{i}}^{\infty}\frac{dz^{\prime}%
}{(1+z^{\prime})E(z^{\prime})}-\int_{z_{F}}^{\infty}\frac{dz^{\prime}%
}{(1+z^{\prime})E(z^{\prime})}\right] \nonumber\\
&  =H_{0}^{-1}\int_{z_{i}}^{z_{F}}\frac{dz^{\prime}}{(1+z^{\prime}%
)E(z^{\prime})}=t_{L}(z_{F})-t_{L}(z_{i})\text{ ,} \label{age}%
\end{align}
where $t_{L}$ is the lookback time, given by
\[
t_{L}(z)=H_{0}^{-1}\int_{0}^{z}\frac{dz^{\prime}}{(1+z^{\prime})E(z^{\prime}%
)}\text{ .}%
\]
Using (\ref{age}), the observational lookback time $t_{L}^{obs}(z_{i})$ is
\begin{align}
t_{L}^{obs}(z_{i})  &  =t_{L}(z_{F})-t(z_{i})=[t_{0}^{obs}-t(z_{i}%
)]-[t_{0}^{obs}-t_{L}(z_{F})]\nonumber\\
&  =t_{0}^{obs}-t(z_{i})-df\text{ ,} \label{lookobs}%
\end{align}
where $t_{0}^{obs}$ is the estimated age of the universe today and $df$ is the
delay factor,
\[
df\equiv t_{0}^{obs}-t_{L}(z_{F})\ .
\]
We now minimize $\chi_{lbt}^{2}$,
\[
\chi_{lbt}^{2}=\sum_{i=1}^{N}\frac{[t_{L}(z_{i},\vec{p})-t_{L}^{obs}%
(z_{i})]^{2}}{\sigma_{i}^{2}+\sigma_{t_{0}^{obs}}^{2}}\text{ ,}%
\]
where $t_{L}(z_{i},\vec{p})$ is the theoretical value of the lookback time in
$z_{i}$, $\vec{p}$ denotes the theoretical parameters, $t_{L}^{obs}(z_{i})$ is
the corresponding observational value given by (\ref{lookobs}), $\sigma_{i}$
is the uncertainty in the estimated age $t(z_{i})$ of the object at $z_{i}$,
which appears in (\ref{lookobs}) and $\sigma_{t_{0}^{obs}}$ is the uncertainty
in getting $t_{0}^{obs}$. The delay factor $df$ appears because of our
ignorance about the redshift formation $z_{F}$ of the object and has to be
adjusted. Note, however, that the theoretical lookback time does not depend on
this parameter, and we can marginalize over it.

In \cite{age35} and \cite{age32} the ages of 35 and 32 red galaxies are
respectively given. For the age of the universe one can adopt $t_{0}%
^{obs}=13.75\pm0.11Gyr$ \cite{wmap7yr}. Although this estimate for
$t_{0}^{obs}$ has been obtained assuming a $\Lambda CDM$ universe, it does not
introduce systematical errors in the calculation: any systematical error
eventually introduced here would be compensated by the adjust of $df$, in
(\ref{lookobs}). On the other hand, such an estimate is in perfect agreement
with other estimates, which are independent of the cosmological model, as for
example $t_{0}^{obs}=12.6_{-2.4}^{+3.4}Gyr$, obtained from globular cluster
ages \cite{krauss} and $t_{0}^{obs}=12.5\pm3.0Gyr$, obtained from
radioisotopes studies \cite{cayrel}.

The WMAP distance information used by the WMAP colaboration includes the
\textquotedblleft shift parameter" $R$, the \textquotedblleft acoustic scale"
$l_{A}$ and the redshift of decoupling $z_{\ast}$. These quantities are very
weakly model dependent \cite{liR}. $R$ and $l_{A}$ are given by%
\[
R=\sqrt{\Omega_{m0}}H_{0}r\left(  z_{\ast}\right)
\]
and%
\[
l_{A}=\pi\frac{r\left(  z_{\ast}\right)  }{r_{s}\left(  z_{\ast}\right)
}\text{ ,}%
\]
where $r\left(  z_{\ast}\right)  $ is the comoving distance to $z_{\ast}$ and
$r_{s}\left(  z_{\ast}\right)  $ is the comoving sound horizon at $z_{\ast}$.
For a flat universe, $r\left(  z_{\ast}\right)  $ and $r_{s}\left(  z_{\ast
}\right)  $ are given by%
\[
r\left(  z_{\ast}\right)  =\frac{1}{H_{0}}\int_{0}^{z_{\ast}}\frac{dz}{E(z)}%
\]
and%
\[
r_{s}\left(  z_{\ast}\right)  =\frac{1}{H_{0}}\int_{0}^{z_{\ast}}\frac
{dz}{E(z)\sqrt{3\left(  1+\bar{R}_{b}/(1+z)\right)  }}\text{ ,}%
\]
where $\bar{R}_{b}=3\Omega_{b0}/\left(  4\Omega_{\gamma0}\right)  $. For the
redshift of decoupling $z_{\ast}$ we use the fitting function proposed by Hu
and Sugiyama \cite{sugiyama}:%
\[
z_{\ast}=1048\left[  1+0.00124\left(  \Omega_{b0}h^{2}\right)  ^{-0.738}%
\right]  \left[  1+g_{1}\left(  \Omega_{m0}h^{2}\right)  ^{g_{2}}\right]
\text{ ,}%
\]
where%
\[
g_{1}=\frac{0.0783\left(  \Omega_{b0}h^{2}\right)  ^{-0.238}}{1+39.5\left(
\Omega_{b0}h^{2}\right)  ^{0.763}}%
\]
and%
\[
g_{2}=\frac{0.560}{1+21.1\left(  \Omega_{b0}h^{2}\right)  ^{1.81}}\text{ .}%
\]
Thus we add to $\chi^{2}$ the term%
\[
\chi_{CMB}^{2}=%
{\displaystyle\sum\limits_{ij}}
\left(  x_{i}^{th}-x_{i}^{data}\right)  \left(  C^{-1}\right)  _{ij}\left(
x_{j}^{th}-x_{j}^{data}\right)  \text{ ,}%
\]
where $x=\left(  l_{A},R,z_{\ast}\right)  $ is the parameter vector and
$\left(  C^{-1}\right)  _{ij}$ is the inverse covariance matrix for the
seven-year WMAP distance information \cite{wmap7ykomatsu}.

Baryonic Acoustic Oscilations (BAO) are described in terms of the parameter
\[
A=\sqrt{\Omega_{M}}E(z_{BAO})^{-1/3}\left[  \frac{1}{z_{BAO}}\int_{0}%
^{z_{BAO}}\frac{dz^{\prime}}{E(z^{\prime})}\right]  ^{2/3}\text{ ,}%
\]
where $z_{BAO}=0.35$. It has been estimated that $A_{obs}=0.493\pm0.017$
\cite{BAO1}. We thus add to $\chi^{2}$ the term
\[
\chi_{BAO}^{2}=\frac{\left(  A-A_{obs}\right)  ^{2}}{\sigma_{A}^{2}}\ \text{.}%
\]

The BAO distance ratio $r_{BAO}\equiv D_{V}\left(  z=0.35\right)
/D_{V}\left(  z=0.20\right)  =1.812\pm0.060$, estimated from the joint
analysis of the 2dFGRS (Two Degree Field Galaxy Redshift Survey) and SDSS
(Sloan Digital Sky Survey)\ data \cite{BAO2}, has also been included. It was
demonstrated in \cite{BAO2} that this quantity is weakly model dependent. The
quantity $D_{V}\left(  z_{BAO}\right)  $ is given by%
\[
D_{V}\left(  z_{BAO}\right)  =c\left[  \frac{z_{BAO}}{H\left(  z_{BAO}\right)
}\left(  \int_{0}^{z_{BAO}}\frac{dz^{\prime}}{H\left(  z^{\prime}\right)
}\right)  ^{2}\right]  ^{1/3}\text{ .}%
\]
So we have the contribution%
\[
\chi_{r_{BAO}}^{2}=\frac{\left(  r_{BAO}-r_{BAO}^{obs}\right)  ^{2}}%
{\sigma_{r_{BAO}}^{2}}\text{ .}%
\]

Finally, we add the 397 supernovae data from Constitution compilation
\cite{constitution}. Defining the distance modulus
\[
\mu(z)=5log_{10}\left[  c(1+z)\int_{0}^{z}\frac{dz^{\prime}}{E(z^{\prime}%
)}\right]  +25-5log_{10}H_{0}\text{ ,}%
\]
we have the contribution
\[
\chi_{SN}^{2}=\sum_{j=1}^{397}\frac{[\mu(z_{j})-\mu_{obs}(z_{j})]^{2}}%
{\sigma_{j}^{2}}\text{ .}%
\]
Using the expression $\chi^{2}=\chi_{lbt}^{2}+\chi_{CMB}^{2}+\chi_{BAO}%
^{2}+\chi_{r_{BAO}}^{2}+\chi_{SN}^{2}$, the likelihood function is given by%
\[
\mathcal{L}(\delta,c,h,\Omega_{b0},\Omega_{\phi_{0}})\propto exp[-\frac
{\chi^{2}(\delta,c,h,\Omega_{b0},\Omega_{\phi_{0}})}{2}]\ \text{.}%
\]

\subsection{Quintessence field}

In table 1 we present the values of the individual best fit parameters, with
respective $1\sigma$, $2\sigma$ and $3\sigma$ confidence intervals.

\begin{center}
\textbf{Table 1}: Values of the holographic quintessence model parameters from
lookback time, CMB, BAO and SNe Ia. In the last line, $\chi_{\min}^{2}/dof$ is
the minimum $\chi^{2}$ per degree of freedom.

\bigskip%

\begin{tabular}
[c]{|l|l|}\hline
$\delta M_{Pl}$ & $-0.170_{-0.072-0.123-0.225}^{+0.067+0.148.+0.187}$\\\hline
$c$ & $0.891_{-0.016-0.028-0.035}^{+0.048+0.134+0.234}$\\\hline
$\Omega_{\phi0}$ & $0.7733_{-0.0087-0.0217-0.0353}^{+0.0092+0.0162+0.0264}%
$\\\hline
$\Omega_{b0}$ & $0.0450_{-0.0028-0.0048-0.0066}^{+0.0026+0.0061+0.0089}%
$\\\hline
$h$ & $0.687\pm0.013\pm0.026\pm0.039$\\\hline
$\chi_{\min}^{2}/dof$ & $1.185$\\\hline
\end{tabular}

\bigskip
\end{center}

\begin{figure}[ptb]
\begin{center}
\includegraphics[width=6.23cm,height=5.0cm]{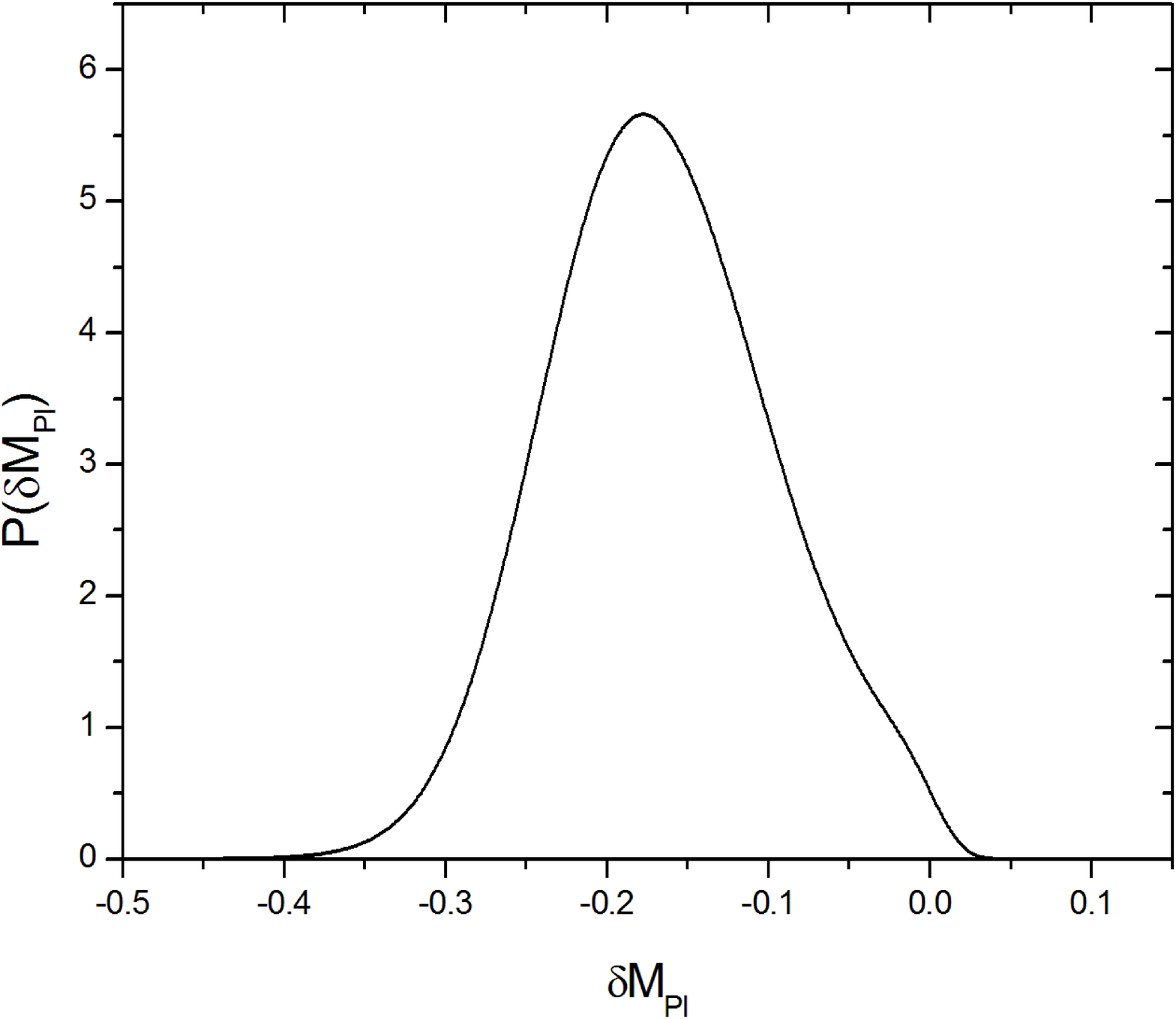}
\ \includegraphics[width=6.23cm,height=5.0cm]{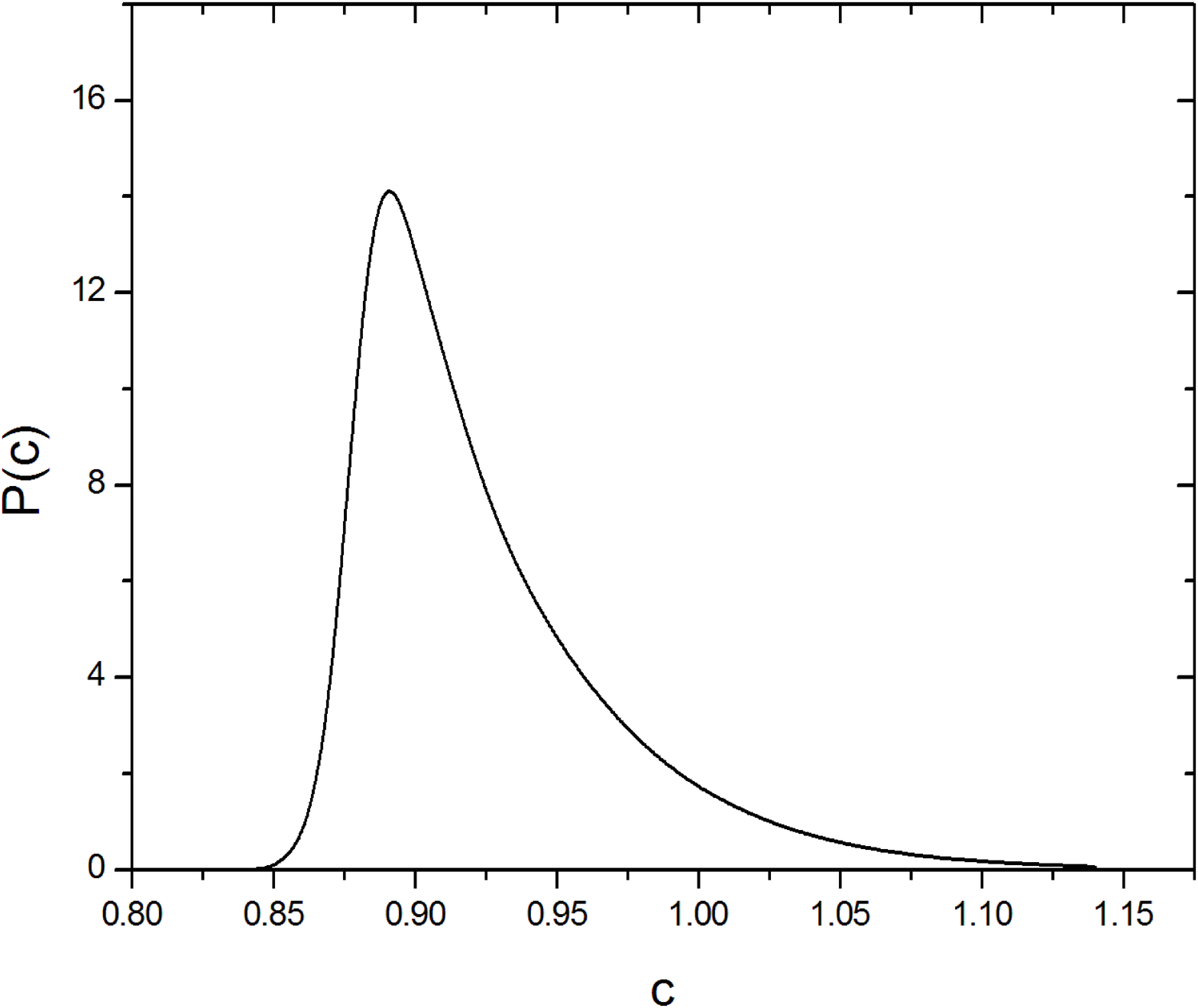}\newline
\end{center}
\caption{Probability distribuctions of the coupling constant $\delta$ (left
panel) and of the parameter $c$ (right panel) of the holographic quintessence
model.}%
\label{distributions_esc}%
\end{figure}

\begin{figure}[ptb]
\begin{center}
\includegraphics[width=5.66cm,height=4.55cm]{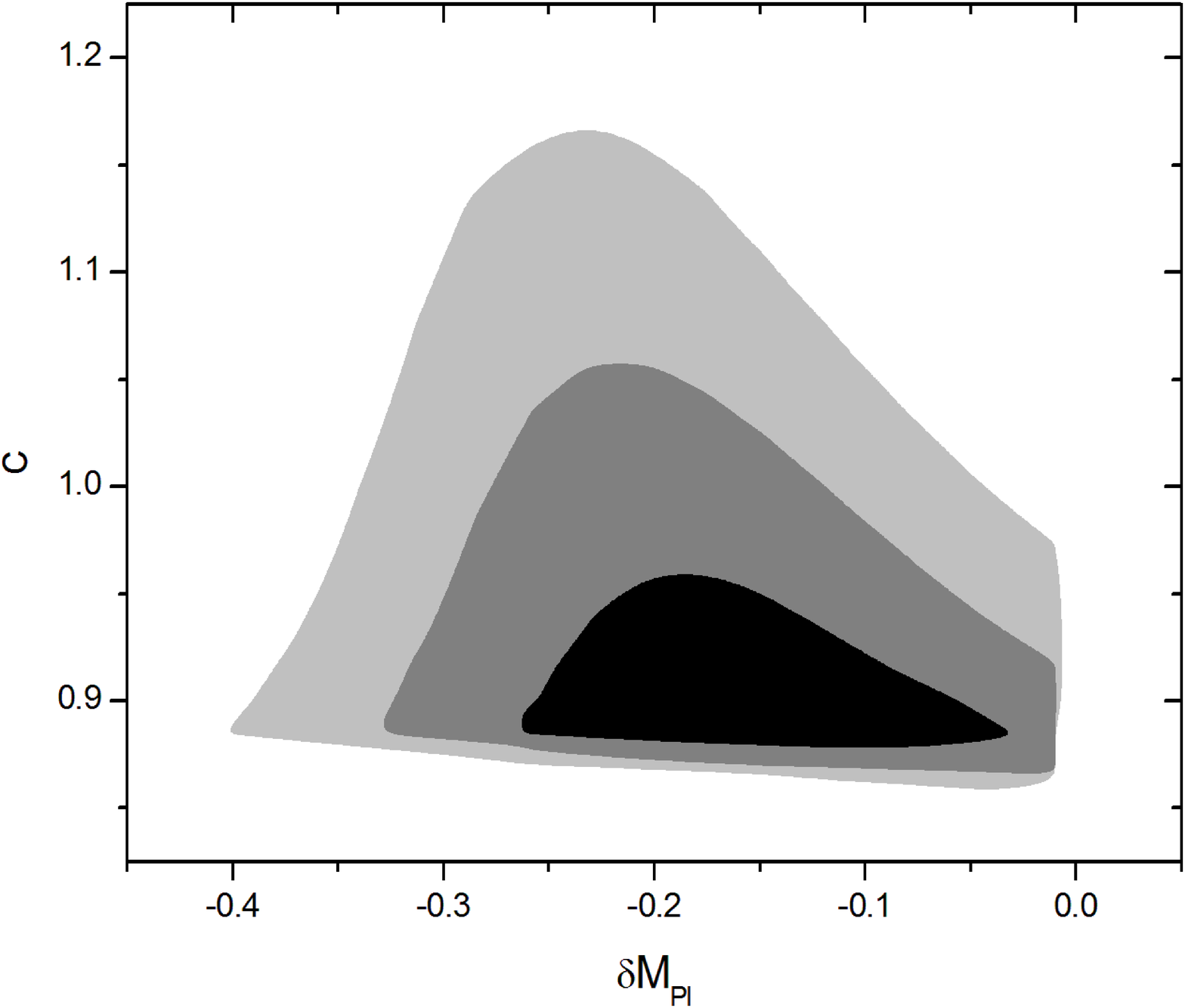}\includegraphics[width=5.66cm,height=4.55cm]{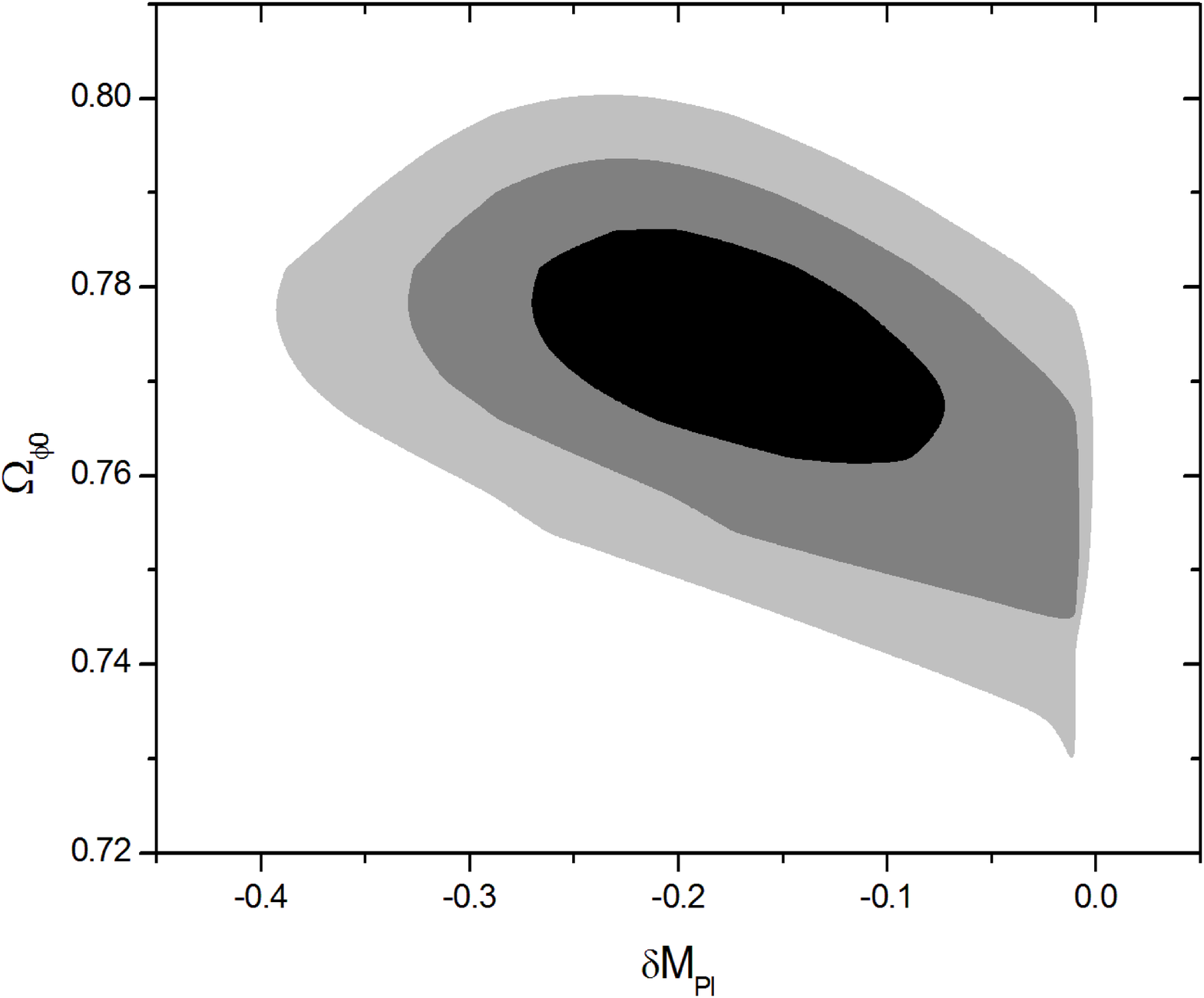}\includegraphics[width=5.66cm,height=4.55cm]{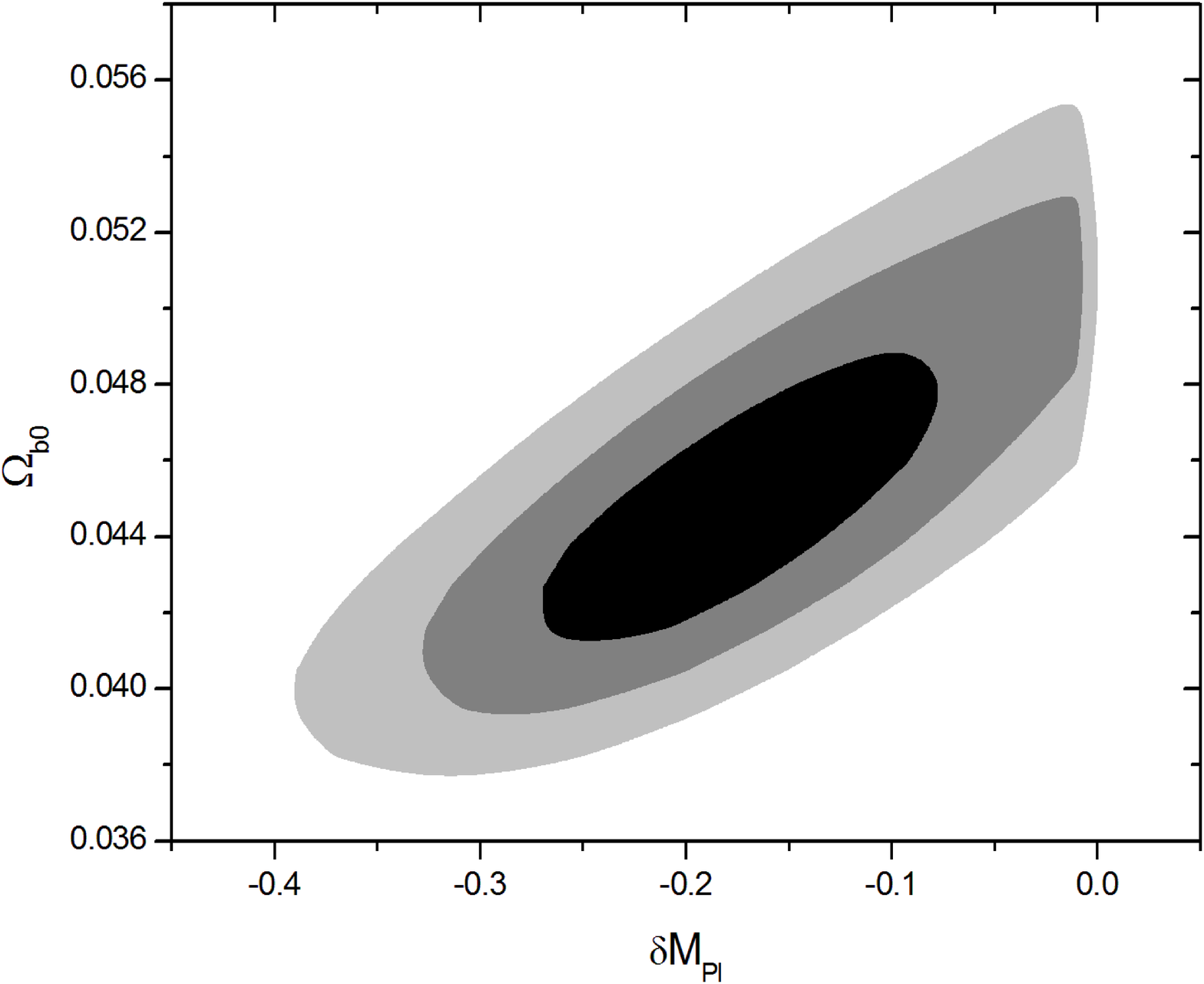}
\includegraphics[width=5.66cm,height=4.55cm]{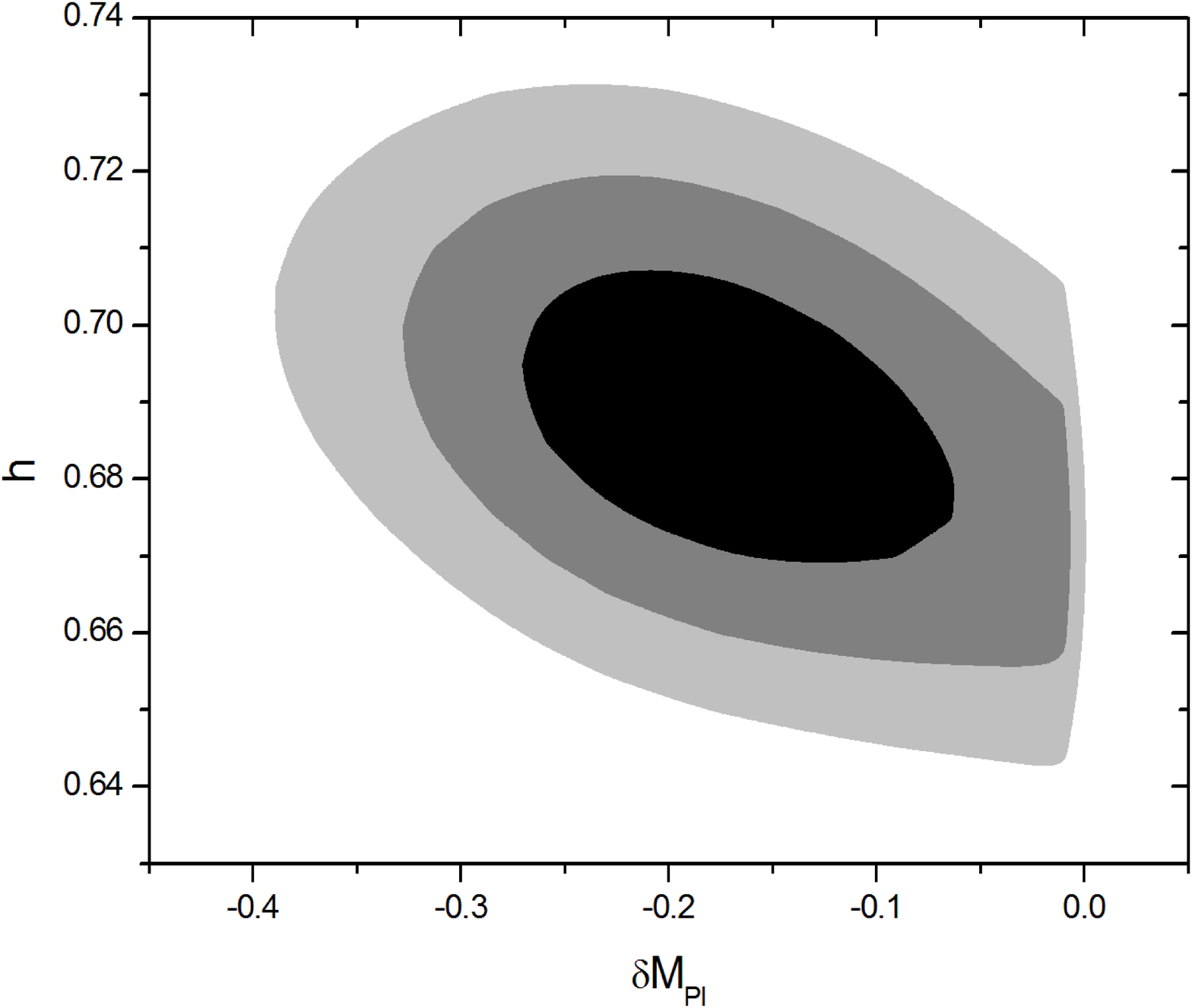}\includegraphics[width=5.66cm,height=4.55cm]{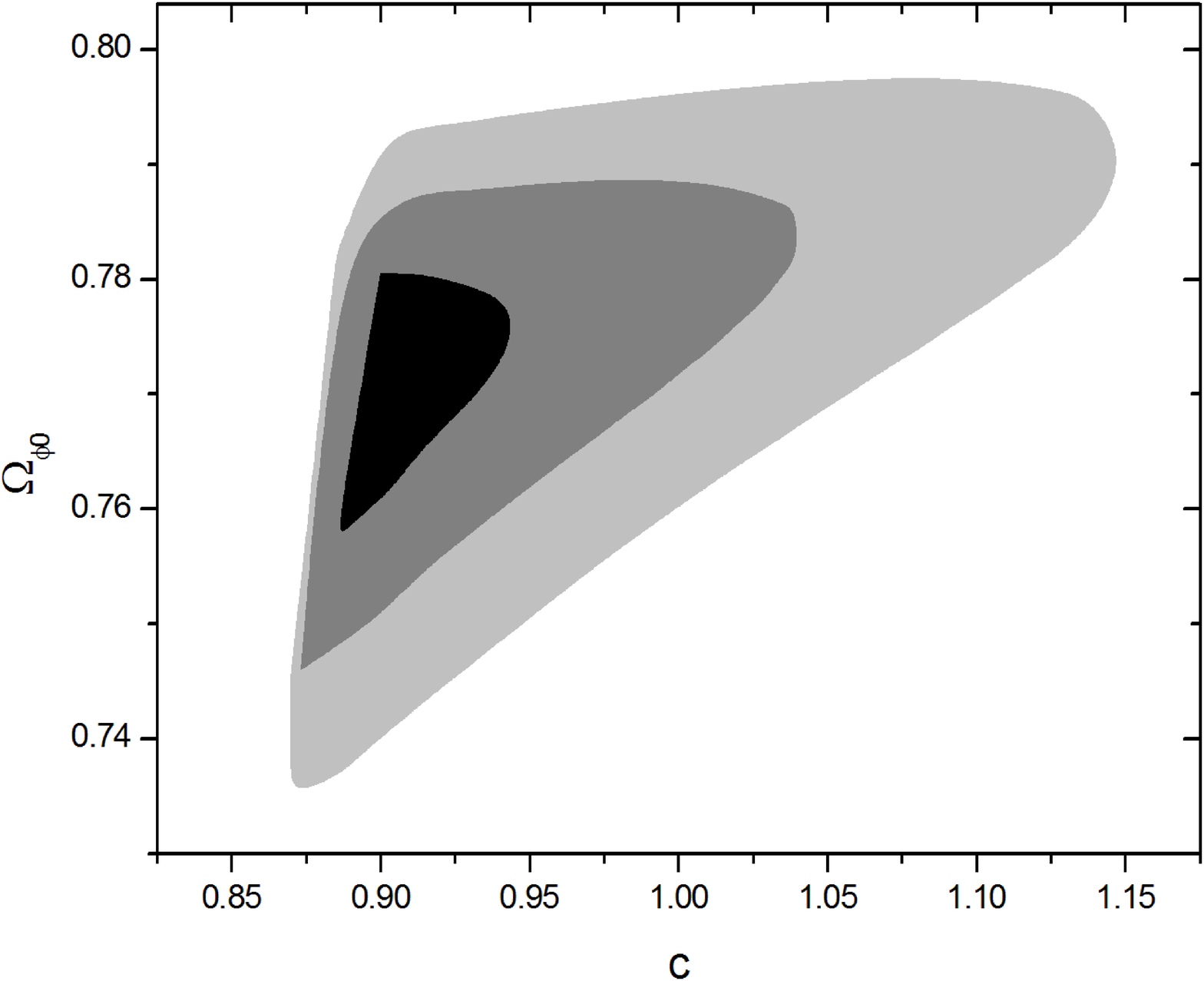}\includegraphics[width=5.66cm,height=4.55cm]{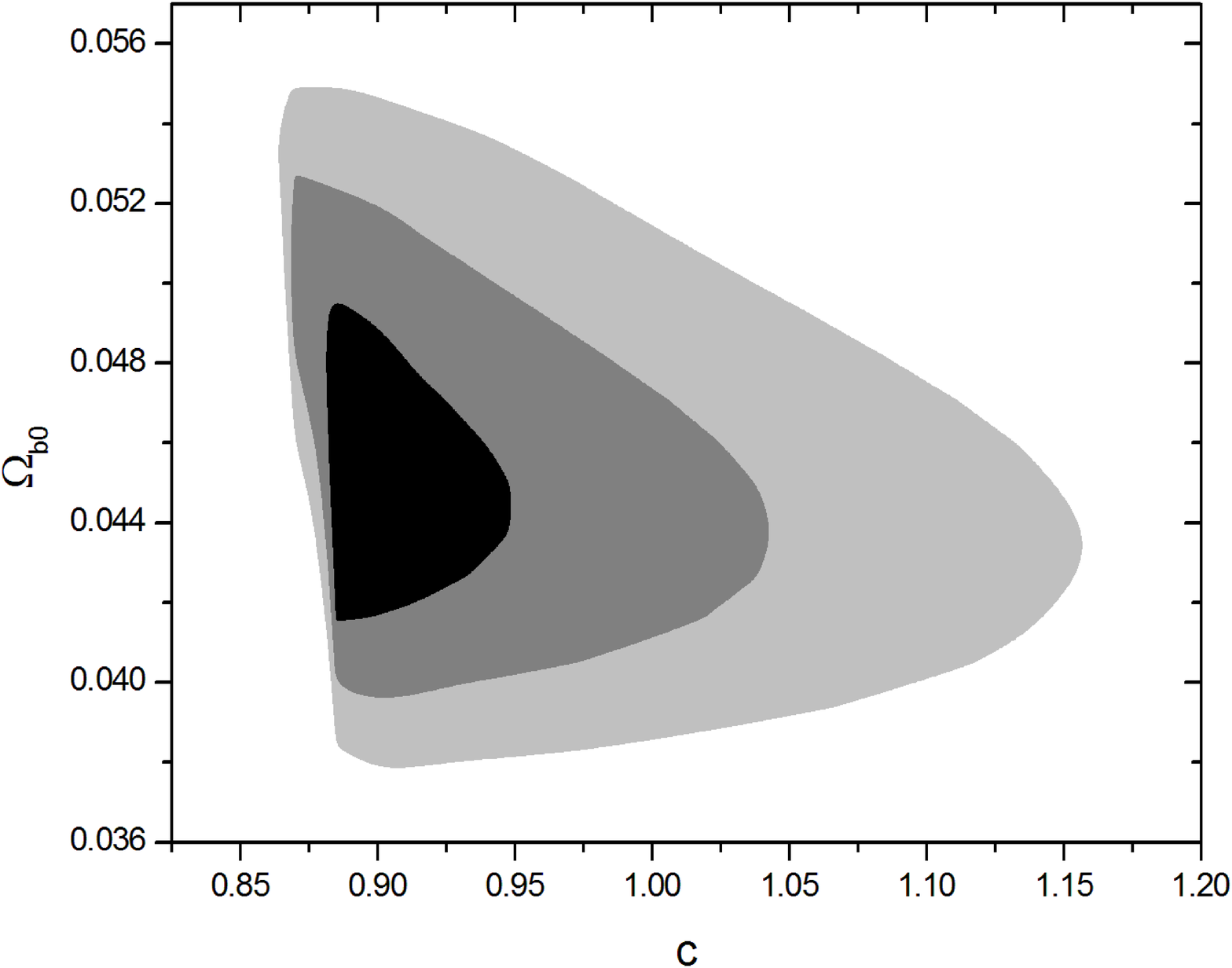}
\end{center}
\caption{Two parameters confidence regions of $1\sigma$, $2\sigma$ and
$3\sigma$ of the holographic quintessence model.}%
\label{bidimensionals_esc}%
\end{figure}

Figure \ref{distributions_esc} shows the marginalized probability
distributions for $\delta$ and $c$. The coupling constant $\delta$ is non
vanishing at $2\sigma$ confidence level. Figure \ref{bidimensionals_esc} shows
some joint confidence regions of two parameters. Notice the effect of the
condition\textbf{ }$\omega_{\phi}\leq1$\textbf{ }on the positive tail of the
probability distribuction of $\delta$ and also in the confidence regions of
$\delta$ with other parameters.

\newpage

\subsection{Tachyon scalar field}

In table 2 we present the values of the individual best fit parameters, with
respective $1\sigma$, $2\sigma$ and $3\sigma$ confidence intervals.

\begin{center}
\textbf{Table 2}: Values of the holographic tachyon model parameters from
lookback time, CMB, BAO and SNe Ia. In the last line, $\chi_{\min}^{2}/dof$ is
the minimum $\chi^{2}$ per degree of freedom.

\bigskip%

\begin{tabular}
[c]{|l|l|}\hline
$\frac{\delta}{H_{0}}$ & $-0.201_{-0.059-0.138-0.249}^{+0.063+0.117+0.176}%
$\\\hline
$c$ & $0.868_{-0.020-0.030-0.038}^{+0.059+0.165+0.235}$\\\hline
$\Omega_{\phi0}$ & $0.724_{-0.012-0.026-0.040}^{+0.013+0.025+0.036}$\\\hline
$\Omega_{b0}$ & $0.0480\pm0.0021\pm0.0041\pm0.0062$\\\hline
$h$ & $0.669\pm0.012\pm0.025\pm0.037$\\\hline
$\chi_{\min}^{2}/dof$ & $1.148$\\\hline
\end{tabular}

\bigskip
\end{center}

\begin{figure}[ptb]
\begin{center}
\includegraphics[width=6.23cm,height=5.0cm]{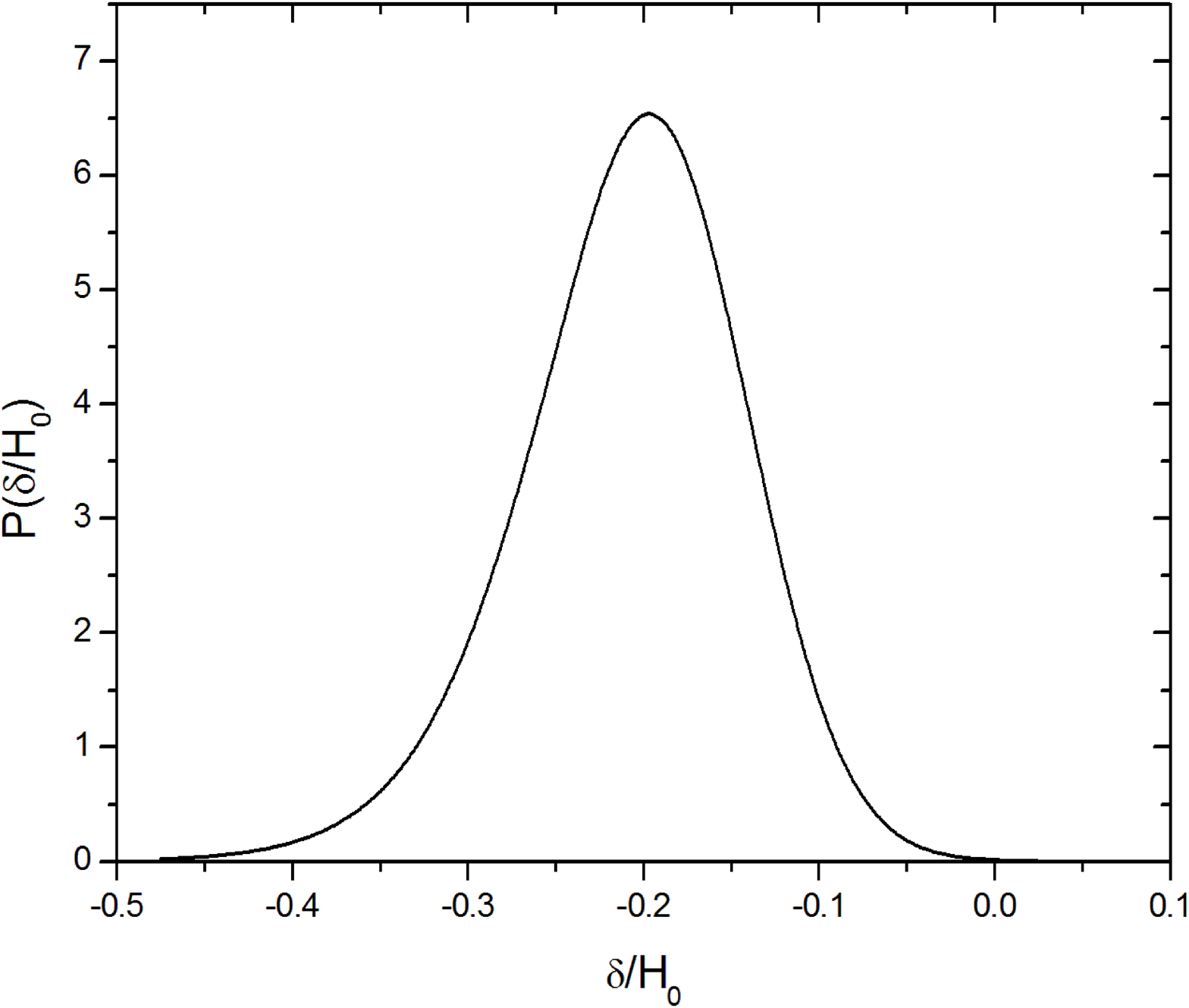}
\ \includegraphics[width=6.23cm,height=5.0cm]{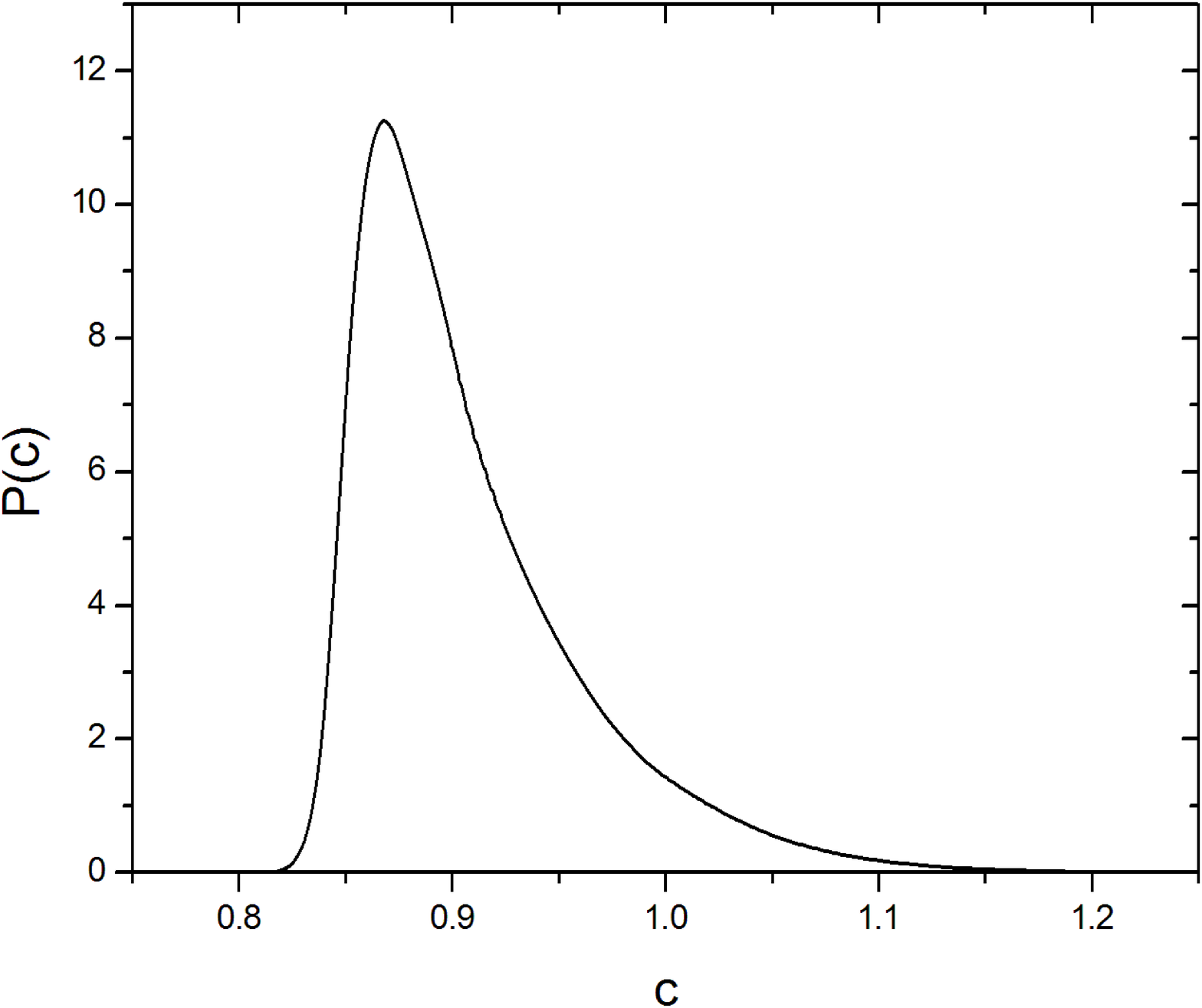}
\end{center}
\caption{Probability distribuctions of the coupling constant $\delta$ (left
panel) and of the parameter $c$ (right panel) of the holographic tachyon
model.}%
\label{distributions_tac}%
\end{figure}

\begin{figure}[ptb]
\begin{center}
\includegraphics[width=5.66cm,height=4.55cm]{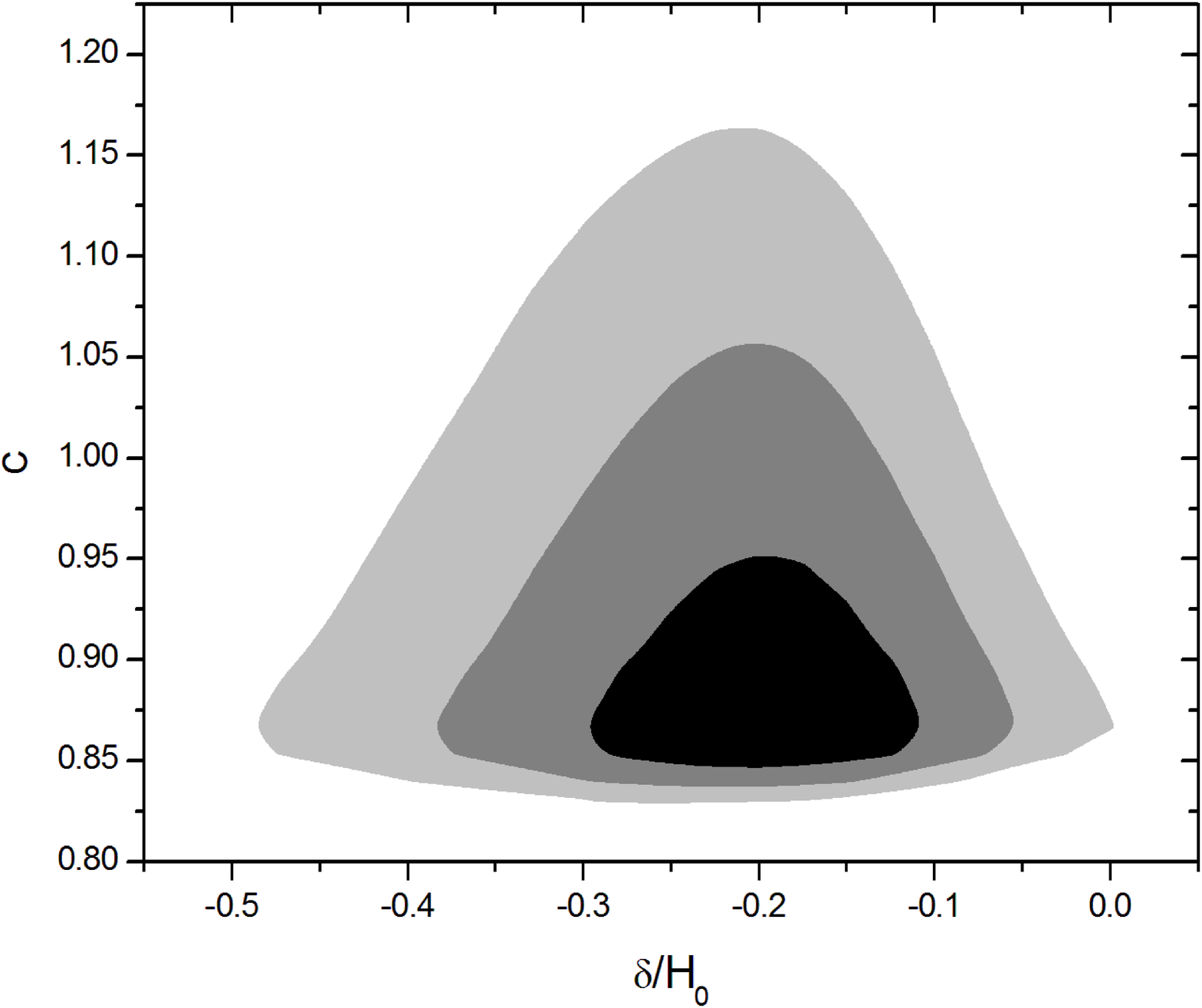}\includegraphics[width=5.66cm,height=4.55cm]{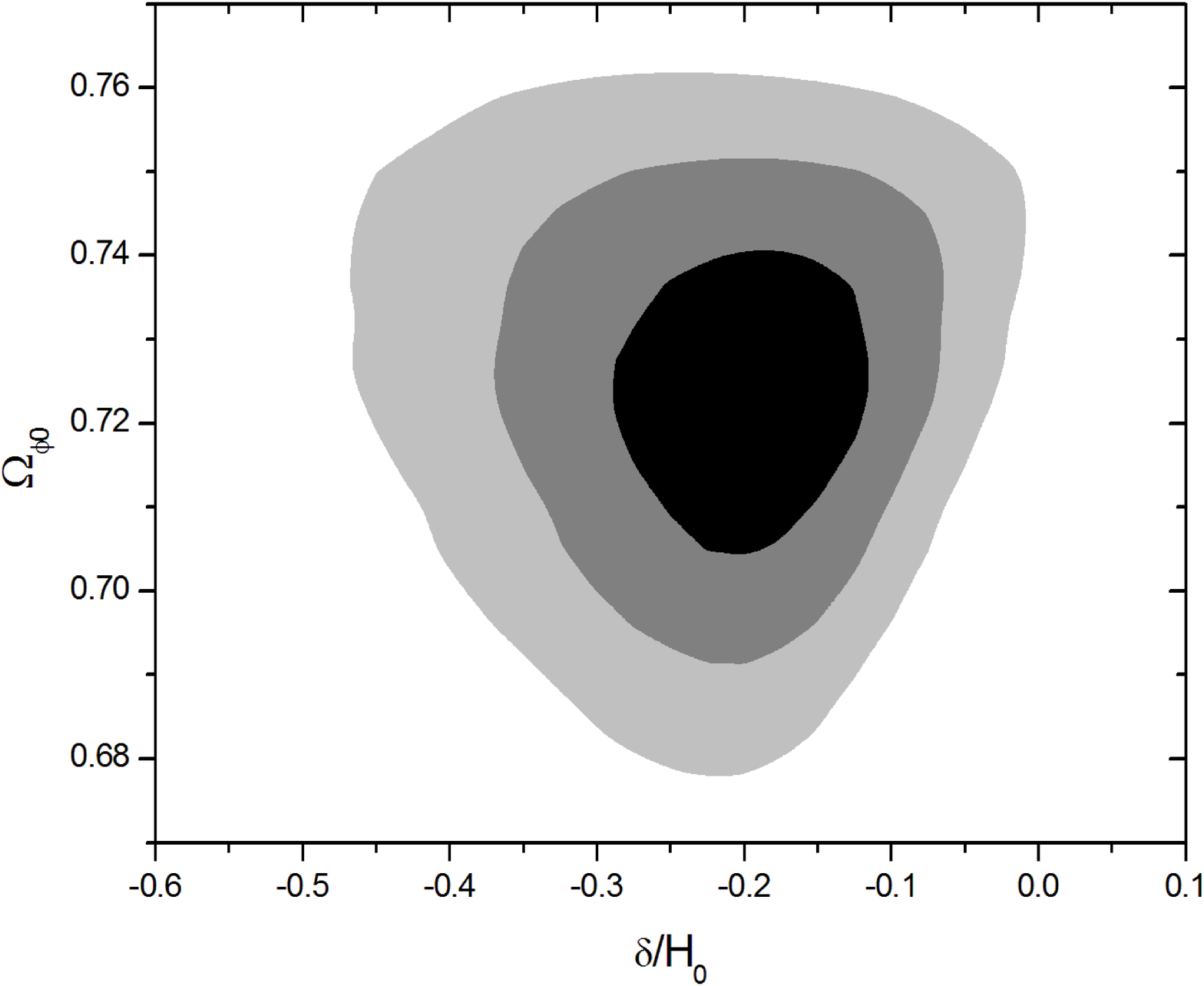}\includegraphics[width=5.66cm,height=4.55cm]{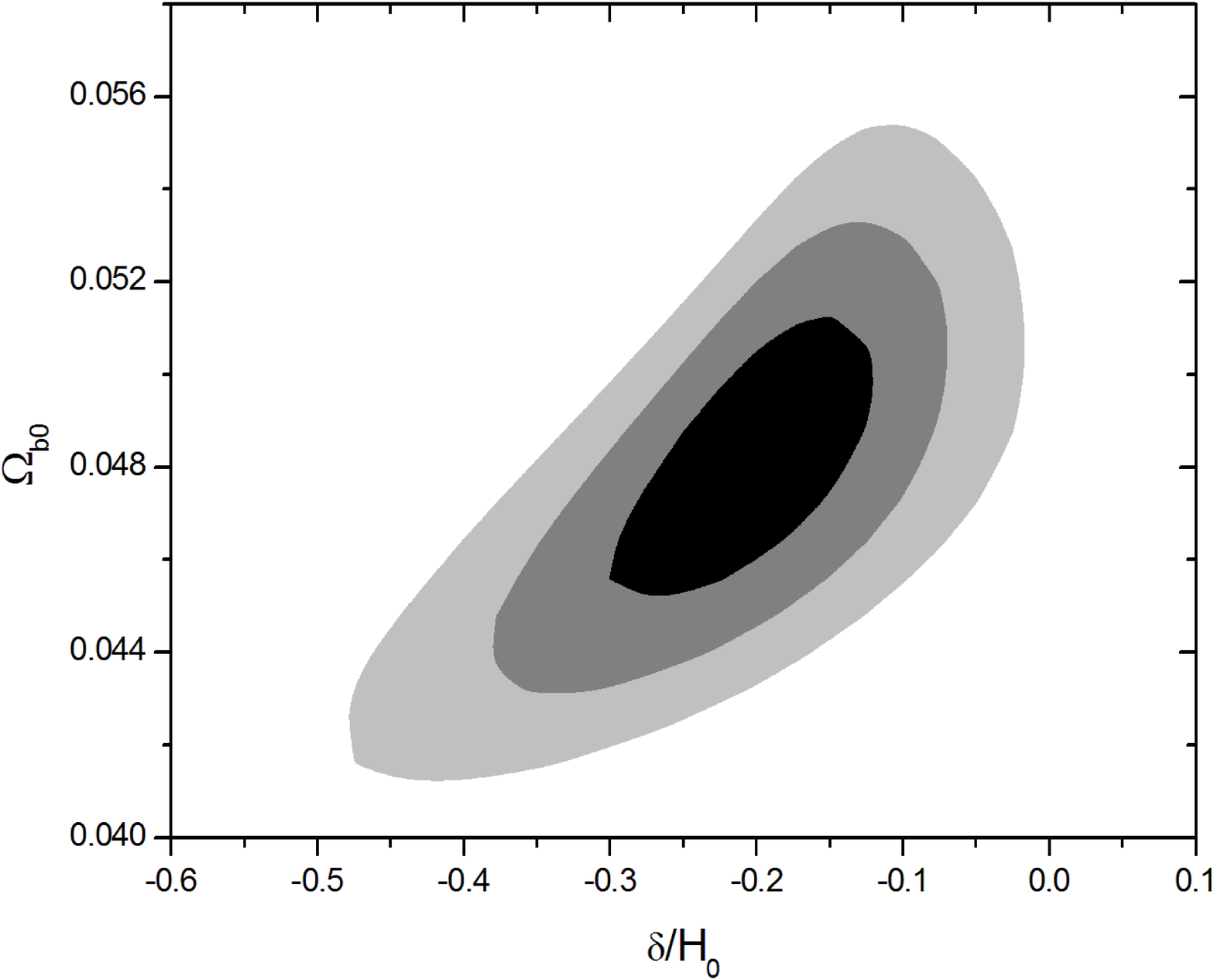}
\ \includegraphics[width=5.66cm,height=4.55cm]{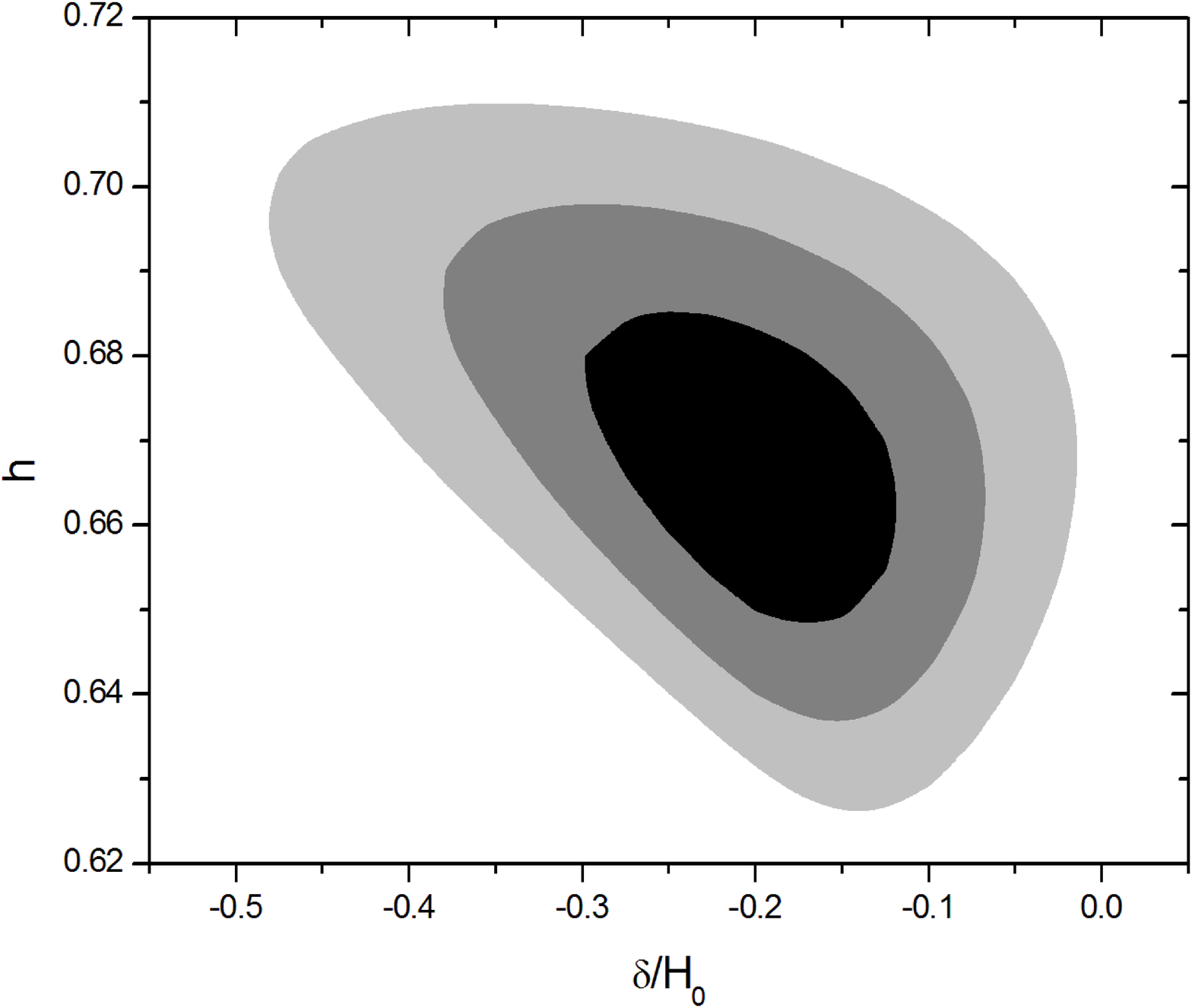}\includegraphics[width=5.66cm,height=4.55cm]{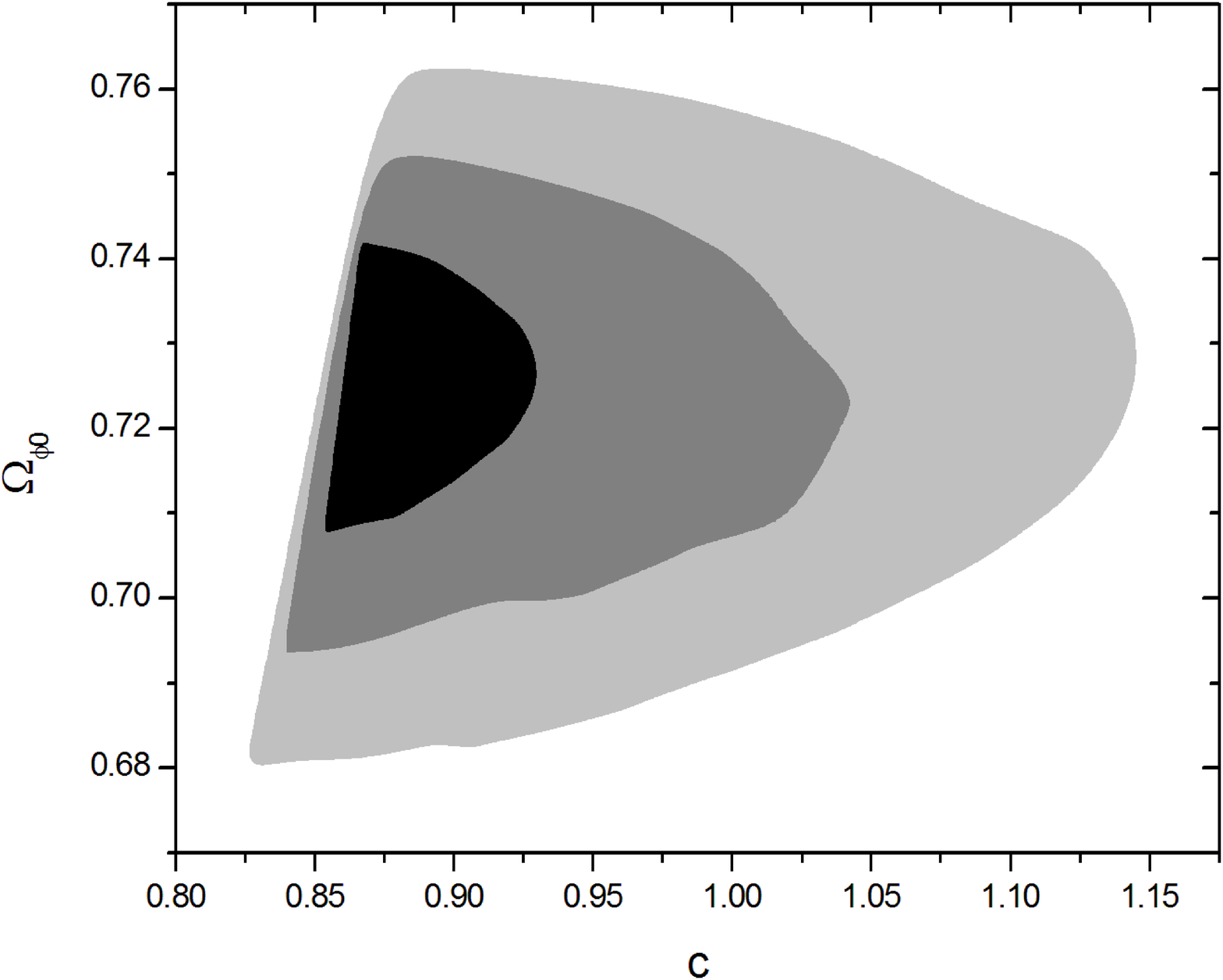}\includegraphics[width=5.66cm,height=4.55cm]{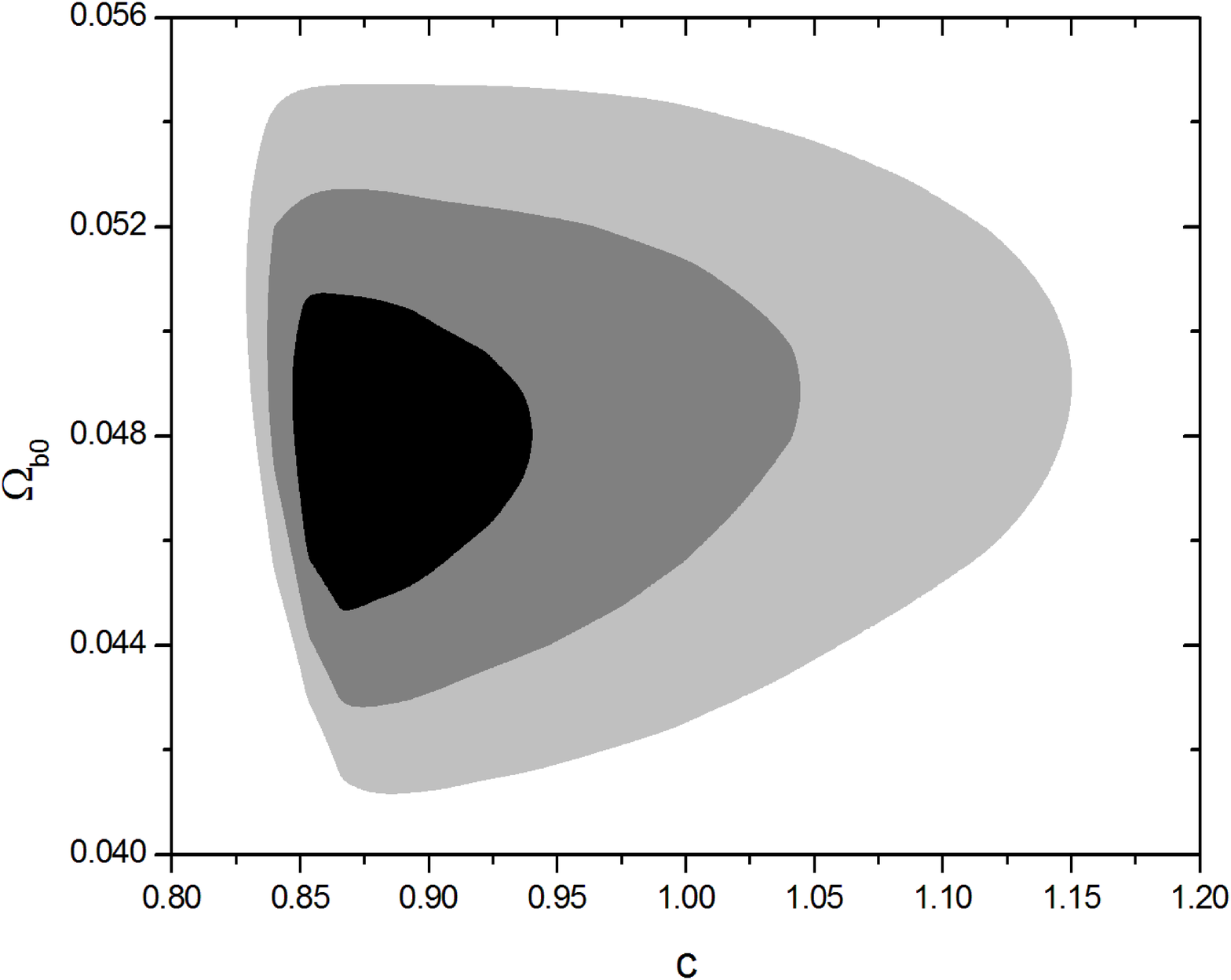}
\end{center}
\caption{Two parameters confidence regions of $1\sigma$, $2\sigma$ and
$3\sigma$ of the holographic tachyon model.}%
\label{bidimensionals_tac}%
\end{figure}

Figure \ref{distributions_tac} shows the marginalized probability
distributions for $\delta$ and $c$. The coupling constant $\delta$ is non
vanishing at more than $3\sigma$ confidence level. So for this model we
obtained strong evidence for interaction. Figure \ref{bidimensionals_tac}
shows some confidence regions of two parameters for this
model.\newpage

\section{Discussion and conclusions}

The minimum $\chi^{2}$ per degree of freedom values indicate that both models
fit well the observational data. Moreover, we see that the tachyon model is a
bit more favored by the observational data than the quintessence one.\textbf{
}The dimensionless coupling constants, $\delta M_{Pl}$ for the quintessence
field and $\frac{\delta}{H_{0}}$ for the tachyon, agree at $1\sigma$ level and
for both models we obtained significative evidence for a non vanishing
interaction in the dark sector. Furthermore, both the results implies in dark
energy decaying into dark matter, alleviating the coincidence problem.\textbf{
}These results are consistent with previous ones, as for example those
obtained in \cite{sandro2}, where an interacting tachyonic dark energy model
with a power law potential was assumed and the results were consistent with
the interaction at $90\%$ confidence level. We must also mention that in
\cite{sandro} and \cite{virial} evidence for interaction was found using
completely different models and data sets. So those results combined with the
present ones furnish sensible evidence in favor of an interaction in the dark
sector of the universe.

The results obtained for $c$, $\Omega_{b0}$ and $h$ for the two models also
agree at $1\sigma$ confidence level. The values obtained for $\Omega_{\phi0}$
agree only at $2\sigma$ level. For the quintessence model, $\Omega_{\phi0}$ is
almost superestimated, corresponding to a matter relative density today
$\Omega_{m0}=0.231\pm0.010$. This value agrees at inferior limit of $1\sigma$
with the cosmological model independent estimative $\Omega_{Mobs}=0.28\pm0.04$
\cite{riess}. For the tachyon model, $\Omega_{\phi0}$ corresponds to a matter
relative density today $\Omega_{m0}=0.276_{-0.012}^{+0.013}$, which is in
perfect agreement with that observational estimative. For both models, the
baryionic density and the Hubble parameter today are very reasonable. We have
obtained $\Omega_{b0}h^{2}=0.0212\pm0.0015$ and $\Omega_{b0}h^{2}%
=0.0215\pm0.0012$ from quintessence and tachyon models, respectively. Let's
compare these values, for example, with that obtained from deuterium to
hydrogen abundance ratio \cite{omeara}, $\Omega_{b0}h^{2}=0.0213\pm
0.0013\pm0.0004$, where the errors terms represents the $1\sigma$ errors from
deuterium to hydrogen abundance ratio and the uncertainties in the nuclear
reaction rates, respectively. For the Hubble parameter, we have obtained
$h=0.687\pm0.013$ and $h=0.669\pm0.012$ from the quintessence and tachyon
models, respectively, both in excellent agreement with observational values,
independent of cosmological model, as for example $h_{obs}=0.69\pm0.12$
\cite{age35} and $h_{obs}=0.72\pm0.08$ \cite{key}. We can also compare the
ratios $\Omega_{b0}/\Omega_{m0}=0.248_{-0.023}^{+0.022}$, from the
quintessence model, and $\Omega_{b0}/\Omega_{m0}=0.211\pm0.016$, from the
tachyon model, with the observational value of 2dFGRS colaboration,
$\Omega_{b0}/\Omega_{m0}=0.185\pm0.046$ \cite{2dFGRS}.

We have obtained $c<1$ at $1\sigma$ confidence level for both models. As
already said above, this implies that the equation of state parameter
$\omega_{\phi}$ will not be real for all future times. However, this is not a
very serious problem, because $c$ is compatible with values above unit at
$2\sigma$ confidence level. Moreover, one could say that $c<1$ is only an
effect due to lack of more precise observational data. Anyway, the very simple
models presented here are expected to be only alternatives to an effective
description of a more sophisticated subjacent theory of dark energy. In
principle, nothing guarantees that they will be good descriptions for all
future times.

Figures \ref{bidimensionals_esc} and \ref{bidimensionals_tac} shows some joint
confidence regions of two parameters for both models. In the confidence
regions for $\delta$ versus $c$ and for $c$ versus $\Omega_{\phi0}$, we see
that there is a lower limit on $c$ $\approx0.8$. This also can be seen in the
marginalized probability distributions of $c$, which dies for $c\lesssim0.85$.
This lower limit is explained by the condition $\frac{\sqrt{\Omega_{\phi0}}%
}{c}<1$, necessary for $\omega_{\phi}$ to be real and $\omega_{\phi}>-1$,
discussed above. This limit can be seen more clearly in $c$ versus
$\Omega_{\phi0}$ confidence regions. Moreover, we have $c\simeq\sqrt
{\Omega_{\phi0}}$ for the best fit values of these parameters. This implies
that $\omega_{\phi0}\simeq-1$ and both models approaches $\Lambda CDM$ today.
This is consistent with the fact that, as $\Lambda CDM$ fits all observational
data, then any alternative model must not deviates much from $\Lambda CDM$ for
$z\approx0$. However, for $z>0$, both models are qualitatively different from
$\Lambda CDM$. For the quintessence field we have very different qualitative
behaviours for $\delta<0$, $\delta=0$ and $\delta>0$, as showed in figure
\ref{fig_w_escrad} and discussed in II A. For the tachyon model its behaviour
is qualitatively the same in the three cases, $\omega_{\phi}$ approximates
$-1/3$, as showed in figure \ref{fig_w_taqrad} and discussed in II B.

It is interesting to compare the results obtained in the present work for the
holographic tachyon model with the previous ones, presented in \cite{sandro3},
where a simpler version of the holographic tachyon model, without barions nor
radiation, had been compared with observational data.\textbf{ }The values
obtained here for $c,$ $\Omega_{\phi0}$ and $h$ are the same as before and
with minor incertainty intervals, despite the fact that here we have one more
parameter - $\Omega_{b0}$ -, as we can see by comparing table 2 in the present
work with table 1 in \cite{sandro3}. This is because now it was possible to
use all WMAP distance information $R,$ $l_{A}$ and $z_{\ast}$, as the model
was generalized to include barions and radiation.\textbf{ }However the dark
energy coupling constant $\delta$ now is non vanishing and compatible with
dark energy decaying into dark matter with more than $3\sigma$ confidence
level, whereas in \cite{sandro3} no evidence of interaction had been found.
This can be understood as follows. In low redshifts, the universe is dominated
by dark energy. The dynamics of dark energy in low redshifts is essentially
determined by the equation of state parameter $\omega_{\phi}\left(  z\right)
$, as can be seen in (\ref{eq_mov_holo}). As in this period the data sets were
the same in both the works, so $\omega_{\phi}\left(  z\right)  $ must be
almost the same in this period in both the works. But $\omega_{\phi}\left(
z\right)  $ explicitly depends on the product $\delta\Omega_{\Psi0}$, were
$\delta$ is the coupling constant and $\Omega_{\Psi0}=1-\Omega_{\varphi
0}-\Omega_{b0}-\Omega_{r0}$ is the dark matter relative density, as we can see
in (\ref{wfi_final_tac}) and (\ref{gama_tac}). As in the present work
$\Omega_{\Psi0}$ is less than that in \cite{sandro3} - where $\Omega
_{b0}=\Omega_{r0}\equiv0$ -, it turns out that $\delta$ in the present work is
bigger (in modulus) than that in \cite{sandro3}. However it is important to
point out that in \cite{sandro3} the model was less realistic, as it wasn't
include barions nor radiation. Therefore, the present result favorable to
interaction is more robust.

In summary, combinations of holographic dark energy model and scalar fields
were implemented. It was showed that it is possible to fix the potential of
interacting scalar fields by imposing that the energy density of the scalar
field must match the energy density of the holographic dark energy. A
comparison of the models with recent observational data was made and the
coupling is non vanishing at more than $2\sigma$ for the quintessence field
and at more than $3\sigma$ for the tachyon. In both cases the results are
consistent with dark energy decaying into dark matter, alleviating the
coincidence problem.

\begin{center}

\textbf{Acknowledgements}

\end{center}

This work has been supported by CNPq (Conselho Nacional de Desenvolvimento
Cient\'{\i}fico e Tecnol\'{o}gico) of Brazil.


\begin{thebibliography}{99}                                                                                               %


\bibitem {1}W. Zimdahl and D. Pavon \textit{Phys. Lett. }\textbf{B521} (2001)
133; L. P. Chimento, A. S. Jakubi, D. Pavon and W. Zimdahl \textit{Phys. Rev.
}\textbf{D67} (2003) 083513.

\bibitem {2}J.-H. He and B. Wang \textit{JCAP} \textbf{06 }(2008) 010; C.
Feng, B. Wang, E. Abdalla and R.-K. Su \textit{Phys. Lett. }\textbf{B665}
(2008) 111; J.-H. He, B. Wang and E. Abdalla, Phys. Lett. \textbf{B671}
(2009), 139.

\bibitem {sandro}B. Wang, J. Zang, C.-Y. Lin, E. Abdalla and S. Micheletti,
\textit{Nucl. Phys. }\textbf{B778} (2007) 69.

\bibitem {sandro2}S. Micheletti, E. Abdalla and B. Wang, \textit{Phys. Rev.
}\textbf{D79 }(2009) 123506.

\bibitem {5}B. Gumjudpai, T. Naskar, M. Sami and S. Tsujikawa \textit{JCAP}
\textbf{06} (2005) 007; B. Wang, Y.-G. Gong and E. Abdalla, \textit{Phys.
Lett.} \textbf{B624} (2005) 141; M. R. Setare, \textit{Phys. Lett.}
\textbf{B642} (2006) 1; \textit{Eur. Phys. J. }\textbf{C50} (2007) 991; idem,
\textit{Phys. Lett. }\textbf{B654 }(2007) 1; E. Abdalla and B. Wang
\textit{Phys. Lett. }\textbf{B651} (2007) 89; R. Rosenfeld \textit{Phys. Rev.
}\textbf{D75} (2007) 083509; M. Quartin, M. O. Calvao, S. E. Joras, R. R. R.
Reis and I. Waga \textit{JCAP} \textbf{05} (2008) 007; Q. Wu, Y. Gong, A. Wang
and J.S. Alcaniz \textit{Phys. Lett. }\textbf{B659} (2008) 34; M.R. Setare and
E. C. Vagenas \textit{Phys. Lett. }\textbf{B666} (2008) 111; M. Jamil, M. A.
Rashid \textit{Eur. Phys. J. }\textbf{C56} (2008) 429; \textit{Eur. Phys. J.
}\textbf{C58} (2008) 111; M.R. Setare and E. C. Vagenas, \textit{Int. J. Mod.
Phys. }\textbf{D18 }(2009) 147; X.-M. Chen, Y.-G. Gong and E. N.
Saridakis,\textit{JCAP }\textbf{04} (2009) 001; Z.-K. Guo, N. Ohta and S.
Tsujikawa, \textit{Phys. Rev.} \textbf{D76} (2007) 023508; O. Bertolami, F.
Gil Pedro and M. Le Delliou, \textit{Phys. Lett. }\textbf{B654} (2007) 165; O.
Bertolami, F.Gil Pedro and M.Le Delliou, \textit{Gen. Rel. Grav. }\textbf{41
}(2009) 2839; L. P. Chimento, \textit{Phys. Rev. }\textbf{D81 }(2010) 043525.

\bibitem {sandro3}S. Micheletti, \textit{JCAP }\textbf{05} (2010) 009.

\bibitem {peebles}P. J. E. Peebles, \textit{Physical Cosmology}, (Princeton U.
Press, 1993).

\bibitem {virial}E. Abdalla, L. R. W. Abramo, L. Sodre Jr. and B. Wang,
\textit{Phys. Lett.} \textbf{B673}, (2009) 107; E. Abdalla, L. R. W. Abramo
and J. C. C. de Souza, \textit{Phys. Rev.} \textbf{D82} (2010) 023508.

\bibitem {bean}R. Bean, E. E. Flanagan, I. Laszlo and M. Trodden \textit{Phys.
Rev. } \textbf{D78} (2008) 123514.

\bibitem {copeland}E. J. Copeland, M. Sami and S. Tsujikawa, \textit{Int. J.
Mod. Phys. }\textbf{D15 }(2006) 1753; M. Sami \textit{Curr. Sci. }\textbf{97
}(2009) 887.

\bibitem {canonico}I. Zlatev, L. Wang and P. J. Steinhardt, \textit{Phys. Rev.
Lett. }\textbf{82} (1999) 896; P. J. Steinhardt, L. Wang and I. Zlatev,
\textit{Phys. Rev. }\textbf{D59} (1999)\textbf{ }123504; L. Amendola
\textit{Phys. Rev. }\textbf{D62} (2000) 043511; R.R. Caldwell and E. V.
Linder, \textit{Phys. Rev. Lett.}\textbf{95} (2005) 141301; R. J. Scherrer and
A. A. Sen, \textit{Phys. Rev.} \textbf{D77} (2008) 083515; A. A. Sen, G. Gupta
and S. Das, \textit{JCAP} \textbf{09} (2009) 027.

\bibitem {sen}A. Sen \textit{JHEP} \textbf{04} (2002) 048; \textit{JHEP}
\textbf{07} (2002) 065; \textit{Mod. Phys. Lett.} \textbf{A17} (2002) 1797.

\bibitem {taquions}T. Padmanabhan \textit{Phys. Rev.} \textbf{D66} (2002)
021301; A. Feinstein \textit{Phys. Rev.} \textbf{D66} (2002) 063511; J. S.
Bagla, H. K. Jassal and T. Padmanabhan \textit{Phys. Rev.} \textbf{D67} (2003)
063504; L. R. W. Abramo and F. Finelli \textit{Phys. Lett.} \textbf{B575}
(2003) 165; R. Herrera, D. Pavon and W. Zimdahl, \textit{Gen. Rel. Grav. }vol.
\textbf{36 }n%
${{}^o}$
\textbf{9 }(2004) 2161; A. Ali, M. Sami and A. A. Sen, \textit{Phys. Rev.
}\textbf{D79} (2009) 123501.

\bibitem {taqholo}J. Zhang, X. Zhang and H. Liu, \textit{Phys. Lett.
}\textbf{B651}, (2007) 84; M. R. Setare, \textit{Phys. Lett. }\textbf{B653},
(2007) 116.

\bibitem {escalarholo}X. Zhang, \textit{Phys. Lett. }\textbf{B648} (2007) 1.

\bibitem {quintomholo}X. Zhang, \textit{Phys. Rev. }\textbf{D74} (2006) 103505.

\bibitem {solarsystem}T. Damour, G. W. Gibbons and C. Gundlach, \textit{Phys.
Rev. Lett. }\textbf{64} (1990) 123.

\bibitem {Li}M. Li, \textit{Phys. Lett. }\textbf{B603}, (2004) 1; Q.-G. Huang
and M. Li, \textit{JCAP} \textbf{08},\textbf{ }(2004) 013.

\bibitem {lookback}S. Capozziello, V. F. Cardone, M. Funaro and S. Andreon
\textit{Phys. Rev.} \textbf{D70} (2004) 123501.

\bibitem {age35}R. Jimenez, L. Verde, T. Treu and D. Stern \textit{Astrophys.
J}. \textbf{593} (2003) 622.

\bibitem {age32}J. Simon, L. Verde and R. Jimenez \textit{Phys. Rev.}
\textbf{D71} (2005) 123001.

\bibitem {wmap7yr}N. Jarosik et. al., \textit{Astrophys. J. Suppl.}
\textbf{192}, (2011) 14.

\textit{WMAP Cosmological Parameters Model/Dataset Matrix homepage,}

\url{http://lambda.gsfc.nasa.gov/product/map/current/best_params.cfm}

\bibitem {krauss}L. M. Krauss \textit{astro-ph/0301012}.

\bibitem {cayrel}R. Cayrel et. al. \textit{Nature} \textbf{409} (2001) 691.

\bibitem {liR}H. Li, J.-Q. Xia, G.-B. Zhao, Z.-H. Fan and X. Zhang,
\textit{Astrophys. J. }\textbf{683 }(2008) L1; Y. Wang and P. Mukherjee
\textit{Phys. Rev.} \textbf{D76} (2007) 103533.

\bibitem {sugiyama}W. Hu and N. Sugiyama, \textit{Astrophys. J. }\textbf{471
}(1996) 542.

\bibitem {wmap7ykomatsu}E. Komatsu, et. al., \textit{Astrophys. J. Suppl.
}\textbf{192 }(2011) 18.

\bibitem {BAO1}B. A. Reid et. al. \textit{0907.1659 [astro-ph.CO]}.

\bibitem {BAO2}W. J. Percival et. al. \textit{Mon. Not. Roy. Astron. Soc.}
\textbf{381 }(2007) 1053.

\bibitem {constitution}M. Hicken et. al. \textit{Astrophys. J}. \textbf{700}
(2009) 1097.

\bibitem {riess}A. G. Riess et. al., \textit{Astrophys. J. }\textbf{659}
(2007) 98.

\bibitem {omeara}J. M. O'Meara, et. al., \textit{Astrophys. J. }\textbf{649}
(2006) L61.

\bibitem {key}W. L. Freedman et. al., \textit{Astrophys. J.} \textbf{553}
(2001) 47.

\bibitem {2dFGRS}S. Cole, et. al., \textit{Mon. Not. Roy. Astron. Soc.
}\textbf{362} (2005) 505
\end{thebibliography}
\end{document}